\documentclass[11pt]{article}

\usepackage[letterpaper,margin=1in]{geometry}

\usepackage[T1]{fontenc}
\usepackage{amsmath}
\usepackage{amsthm}
\usepackage{newtxtext,newtxmath} % supplies AMS symbols; do not also load amssymb
\usepackage{microtype}

\usepackage{graphicx}
\usepackage{booktabs}
\usepackage{multirow}
\usepackage{algorithm}
\usepackage{algpseudocode}
\usepackage{tikz}
\usetikzlibrary{arrows.meta,positioning,calc,fit,backgrounds,decorations.pathreplacing}
\usepackage{adjustbox} % max width=\textwidth: scale wide tikz figures down to fit, never up
\usepackage{caption}
\usepackage{float}

\usepackage{authblk}

\usepackage[colorlinks=true,linkcolor=blue,citecolor=blue,urlcolor=blue]{hyperref}

\newcommand{\Ftwo}{\mathbb{F}_2}
\newcommand{\CNOT}{\textup{CNOT}}

\newcommand{\GL}{\mathrm{GL}}
\newcommand{\sM}{s} % minimal CNOT count (word length) of a matrix
\newcommand{\binmat}[1]{\left[\begin{smallmatrix}#1\end{smallmatrix}\right]}

\begin{document}

\title{\bfseries Noise-Aware Synthesis of Quantum LDPC Encoder Circuits\\ via Two-Sided Hamming Descent}

\author{Aditya Sodhani\thanks{\texttt{sodha005@umn.edu}}}
\author{Keshab K. Parhi\thanks{Corresponding author: \texttt{parhi@umn.edu}}}
\affil{Department of Electrical and Computer Engineering,\\
University of Minnesota, Minneapolis, Minnesota, USA}
\date{}

\maketitle

\begin{abstract}
Quantum low-density parity-check (LDPC) codes are a promising route to fault-tolerant quantum computation, but their use requires efficient preparation of encoded states. Standard encoder constructions generate circuits through fixed algebraic procedures, yet the resulting circuit can contain substantial redundancy. We formulate LDPC encoder preparation as a circuit-resynthesis problem: given the linear-reversible matrix implemented by the encoder's CNOT block, we seek a lower-cost equivalent circuit that can be \textit{routed} efficiently on the target hardware and which mitigates noise. We propose a novel optimization approach referred as \textit{two-sided Hamming descent} and a \textit{noise-aware} optimization pipeline for this task.

Across several families of Calderbank-Shor-Steane (CSS) LDPC encoders, including Bivariate Bicycle, hypergraph-product, and entanglement-assisted codes, the proposed pipeline produces substantially smaller and shallower encoder circuits than the standard constructions and the synthesis baselines considered, cutting gate counts by $53.8\%$ in aggregate across the benchmark and by up to $68\%$ on the Bivariate Bicycle family. The gains remain visible after routing, where the two-qubit depth is reduced by up to $71\%$ and translate into higher-fidelity state preparation under circuit-level noise. On the Bivariate Bicycle family, live-range scheduling further reduces routed preparation failure by up to $13.7\%$ without adding two-qubit gates to the selected circuit. These results indicate that encoder-matrix resynthesis, combined with hardware-calibrated selection and scheduling, is an effective compiler-level tool for preparing quantum LDPC code states.
\end{abstract}

\medskip
\noindent\textbf{Keywords:}\quad quantum LDPC codes, encoder synthesis, linear reversible circuits, CNOT optimization, two-sided Hamming descent, noise-aware optimization, circuit depth, commutation-aware scheduling, Bivariate Bicycle codes, fault-tolerant quantum computing.
\raggedbottom  % natural page bottoms; avoids flushbottom stretching gaps around floats/lists

% ============================================================================
\section{Introduction}
\label{sec:intro}

Quantum low-density parity-check (LDPC) codes are central to practical fault-tolerant quantum computation~\cite{shor1995scheme,breuckmann2021quantum,roffe2019quantum}. Bivariate Bicycle (BB) codes have recently been proposed for superconducting hardware and shown to achieve a high pseudo-threshold under circuit-level noise~\cite{bravyi2024high}. Hypergraph-product (HGP) codes~\cite{tillich2013quantum} provide the first constant-rate family with polynomial distance and underlie subsequent asymptotically good constructions~\cite{hastings2021fiber,panteleev2022asymptotically,leverrier2022quantum}. Entanglement-assisted quasi-cyclic LDPC (EA QC-LDPC) codes extend these constructions to settings with pre-shared entanglement~\cite{kumar2025entanglement,wilde2008optimal,luo2009channel}. Realizing any of these codes begins with the same operational step. An encoder circuit maps unencoded qubits into a valid encoded state, the first step of the fault-tolerant stack. A noisier encoder injects more error into this initial logical state before the error-correction cycle begins, so high-fidelity preparation is a prerequisite for fault-tolerant performance~\cite{bravyi2024high}.

The BB, HGP, and EA QC-LDPC families are all Calderbank-Shor-Steane (CSS)~\cite{steane1996error,calderbank1998quantum} constructions, and their encoder circuits consist of Hadamard gates on a subset of qubits followed by a \CNOT{} subsequence that entangles the qubits into the stabilizer state~\cite{cleve1997efficient,gottesman1998heisenberg,gottesman1997stabilizer,mondal2024quantum}. The standard constructions for this task are the Cleve-Gottesman (CG) reduction~\cite{cleve1997efficient} for standard CSS codes and the Sharma-Kumar-Garani (SKG) construction~\cite{sharma2025encoding} for EA codes. Both construct the encoder circuit from the code's parity-check matrix along a fixed algebraic path (reduced row echelon form with deterministic column ordering). Their fixed elimination paths guarantee a valid encoder through systematic construction, but they are not hardware-aware optimizers. They do not minimize \CNOT{} count, circuit depth, routing overhead, or the noise exposure of the prepared state. The resulting circuits should therefore be optimized before hardware execution~\cite{mondal2024optimization}.

Quantum circuit optimization techniques fall into two broad categories. The first category, \emph{gate-level optimization}, includes peephole optimizers such as Qiskit~O3~\cite{treinish2023qiskit}, TKET~\cite{sivarajah2021t}, the continuous-parameter optimizer of Nam~\emph{et~al.}~\cite{nam2018automated}, and NISQ-oriented passes~\cite{nash2020quantum}, which apply local circuit identities. This category also includes automated identity generation~\cite{pointing2024optimizing}, which discovers such identities algorithmically, and ZX-calculus and Clifford-template pipelines~\cite{kissinger2020reducing,duncan2020graph,bravyi2021clifford}, which rewrite circuits at a higher level of abstraction. The second category, \emph{whole-circuit resynthesis}, discards the input gate sequence and rebuilds the circuit directly from the linear-reversible transformation $M$. The Patel--Markov--Hayes (PMH) algorithm achieves an asymptotically optimal $O(n^2/\log n)$ \CNOT{} count~\cite{markov2008optimal}, and later refinements include greedy, Gaussian-elimination-based, A$^{*}$, beam-search, and learning-based approaches~\cite{de2021gaussian,webster2025heuristic,maslov2023cnot,de2021reducing,romanello2025cnot,sodhani2026optimizing}, as well as connectivity-aware methods such as Steiner-tree synthesis~\cite{kissinger2019cnot} and PermRowCol~\cite{meijer2023dynamic}, and nearest-neighbor-compliant circuit design~\cite{mondal2024optimized}. For these methods, exact synthesis is feasible only for very small instances ($n \lesssim 7$), via SAT/QBF encodings~\cite{schneider2023sat,shaik2024optimal}, breadth-first search over the $\GL(n,\Ftwo)$ Cayley graph~\cite{christensen2025exact}, or meet-in-the-middle database search~\cite{amy2013meet}. Similar approaches have also been proposed for encoder circuit optimization of nonbinary codes in prime dimensions~\cite{sodhani2026encoder}.

For the encoder-preparation problem studied here, both categories leave important gaps. Gate-level optimizers only make local changes, so they can remove local redundancy but not the global overhead that the fixed-order CG/SKG construction~\cite{cleve1997efficient,sharma2025encoding} builds into the encoder. Whole-circuit resynthesis can in principle remove that overhead, but existing methods optimize gate count alone, and the strongest of them are greedy: at each step they choose the gate that removes the most 1s from the matrix they are simplifying (its weight). On the dense matrices that CSS LDPC encoders produce, this weight barely changes from one move to the next, so the search loses its descent direction and recovers only part of the available reduction. This loss of ground is already measurable in the strongest prior synthesizers. The cost-minimization greedy approach of Schaeffer and Perkowski~\cite{schaeffer2014cost} and its systematic development by Goubault de Brugi\`ere~\emph{et~al.}~\cite{de2021gaussian} are fragile precisely on the dense, structured matrices that CSS LDPC encoders produce. At BB-code scale their strict-descent variants stall on a majority of restarts, and the surviving variants leave 8--25\% of the achievable reduction unrealized. The same loss of descent direction also makes them slower for two reasons: stalled restarts produce nothing, and the runs that do converge take more steps. Three further gaps cut across both categories. First, the depth these methods report is the depth of the gates taken in the order the synthesizer outputs them. Commuting gates could be reordered into fewer parallel layers, so this gate-list depth overstates the true parallel cost. Second, the gate-count gains assume a fully connected device on which any qubit can act on any other, whereas real hardware connects each qubit to only a few neighbors, so an LDPC code laid out on that limited graph needs extra gates to bring distant qubits together. Third, count and depth are only proxies for the quantity that ultimately matters, the preparation error of the encoded state under hardware noise. Fewer gates or fewer layers do not by themselves give a lower-error encoder, since the error also depends on how long qubits sit idle and on the extra gates that routing inserts. Prior synthesizers optimize count, and at best depth, but none selects or schedules the encoder to reduce this error directly, so no existing method is noise-aware from end to end.

Our approach is to treat encoder preparation as a single, noise-aware resynthesis problem that closes these gaps together. To this end, we propose a novel \textit{two-sided Hamming descent} optimization approach and develop a complete pipeline for this resynthesis proble. Because exact \CNOT{}-count minimization is NP-hard~\cite{amy2019controlled}, we rebuild each encoder from its target matrix $M$ with a heuristic objective that keeps a useful descent direction on these dense matrices, then carry the result through depth compression, routing, and hardware-calibrated selection and scheduling so that the encoder is optimized for the error it will actually incur on the target device.

\begin{figure}[t]
\centering
\includegraphics[width=\textwidth]{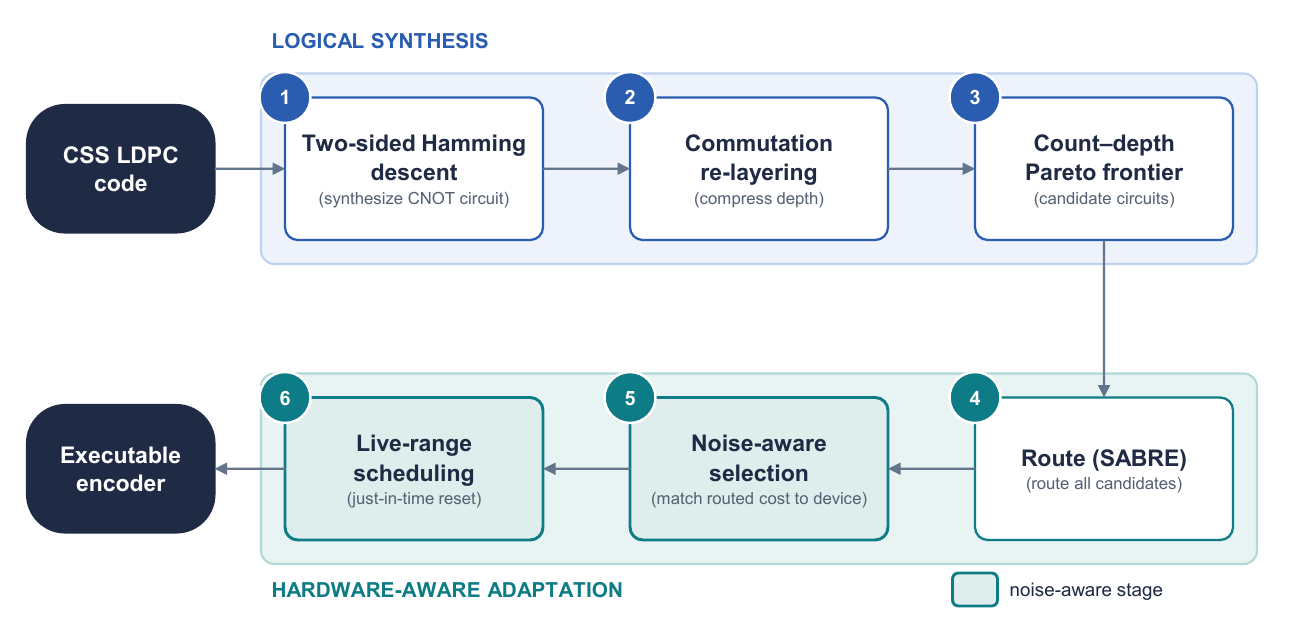}
\caption{Our six-stage noise-aware encoder pipeline.}
\label{fig:pipeline}
\end{figure}

Our central contribution is the six-stage noise-aware encoder pipeline shown in Figure~\ref{fig:pipeline}. The main contributions are as follows.
\begin{enumerate}
  \item \textbf{Two-sided Hamming descent.} We synthesize the encoder matrix $M$ by reducing a residual matrix to the identity using both row and column transvections. Candidate moves are scored by their Hamming-distance reduction and their effect on the circuit depth. Multistart feature explores different synthesis paths, and every resulting circuit is verified against the encoder matrix $M$. This cuts \CNOT{} counts by $53.8\%$ in aggregate and by up to $68\%$ on the Bivariate Bicycle family, beating implementations of the strongest published greedy baselines by $7.6$--$24.6\%$ on every BB code (Sections~\ref{sec:twosided}--\ref{sec:objective}).
  \item \textbf{Commutation re-layering.} We reorder commuting \CNOT{} gates while preserving the precedence constraints between noncommuting gates, place them in disjoint-qubit layers, and verify the result, giving a commutation-aware depth that never exceeds the gate-list ASAP depth. This commutation-aware depth is within $1.05$--$1.17\times$ of a per-circuit lower bound at zero \CNOT{} cost (Section~\ref{sec:relayer}).
  \item \textbf{Count--depth Pareto frontier and routing.} Rather than select a single circuit immediately after synthesis, we retain the non-dominated circuits from the layer-penalty sweep, then re-layer and route each on the code-native coupling graph, so its physical gate count, depth, and idle structure can be evaluated before selection. On the BB-native biplanar architecture this reduces routed two-qubit count by $51.6$--$57.3\%$ and routed two-qubit depth by $55.7$--$71.1\%$ (Sections~\ref{sec:depth-results}--\ref{sec:native-routing}).
  \item \textbf{Noise-aware selection and live-range scheduling.} We select the routed candidate that minimizes a hardware-calibrated preparation cost combining two-qubit-gate exposure and idle exposure. We then schedule it as late as its dependencies permit, with each qubit prepared immediately before its first use. This cuts idle exposure without changing the routed \CNOT{} list or its implemented transformation. It lowers routed preparation failure on $16$ of $17$ codes, by up to $34.7\%$ (Sections~\ref{sec:selectmethod}--\ref{sec:lrmethod}, \ref{sec:pipelinecompare}--\ref{sec:noiseselect}).
\end{enumerate}

\medskip\noindent\textbf{Overview.} Section~\ref{sec:prelim} introduces the encoder-matrix model, transvection notation, and the circuit-cost metrics used throughout the paper, including gate-list ASAP depth and commutation-aware depth. Section~\ref{sec:blts} presents the full optimization pipeline: two-sided Hamming descent, commutation-aware scheduling and verification, noise-aware post-routing selection, and live-range scheduling. Section~\ref{sec:results} evaluates the pipeline on CSS LDPC encoder families, comparing against standard constructions, published greedy methods, and general-purpose compilers, and then studies routing, circuit-level noise, hardware-aware selection, and comparisons. Section~\ref{sec:discussion-conclusion} discusses the mechanisms behind the improvements, the scope of the method, and its limitations.

% ============================================================================
\section{Preliminaries}
\label{sec:prelim}

\subsection{Encoders, transvections, and the word metric}
\label{sec:wordmetric}

\begin{figure}[t]
\centering
\includegraphics[width=0.50\textwidth]{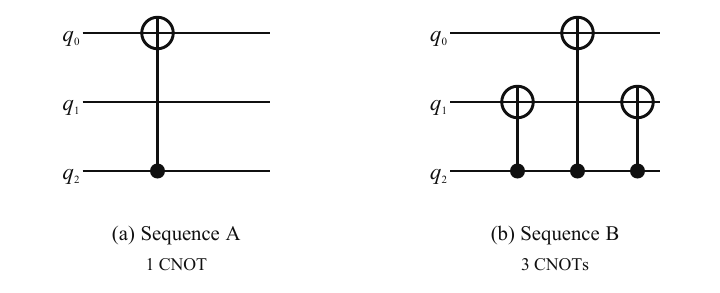}
\caption{The same matrix $M = I + e_0 e_2^{\top}$ on three qubits is realized by \CNOT{} circuits of different length, hence by the same unitary. (a) one gate; (b) three gates. Filled dots are controls, $\oplus$ targets.}
\label{fig:equiv3q}
\end{figure}

\begin{figure}[t]
\centering
\includegraphics[width=0.55\textwidth]{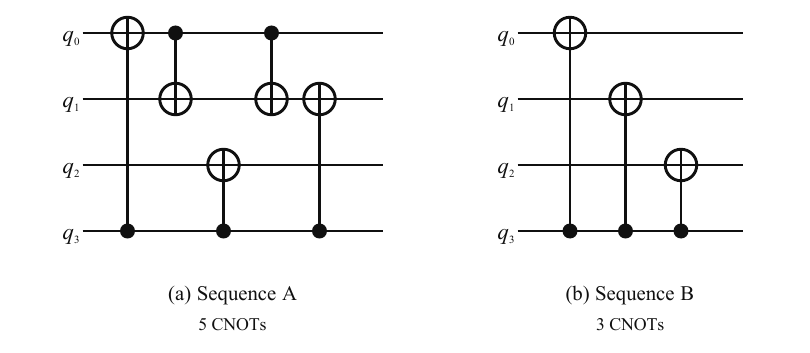}
\caption{The same four-qubit matrix $M$ is implemented by two \CNOT{} circuits of different gate count, so the shorter one replaces the longer. (a) five gates; (b) three gates. Filled dots are controls, $\oplus$ targets.}
\label{fig:equiv4q}
\end{figure}

For a CSS code specified by check matrices $H_X$ and $H_Z$, the Cleve--Gottesman (CG) and Sharma--Kumar--Garani (SKG) constructions produce an encoder of the form $U = C \cdot H^{\otimes S}$, where $H^{\otimes S}$ denotes a layer of Hadamard gates applied to a prescribed subset $S$ of qubits and $C$ is a \CNOT{} circuit. The \CNOT{} block $C$ realizes a linear reversible transformation $M \in \GL(n,\Ftwo)$. Throughout, $Mx$ denotes matrix--vector multiplication over $\Ftwo$, and computational-basis strings are ordered as $q_0 q_1 \cdots q_{n-1}$. The induced \CNOT{} unitary is $U_M = \sum_{x \in \Ftwo^n} |Mx\rangle\langle x|$, so $M$ determines the action of $C$ on every computational-basis state and, by linearity, on the entire Hilbert space. For a gate $\CNOT(j{\to}i)$, the associated transvection is $T_{ij} = I + e_i e_j^{\top}$ (with $i \neq j$), which acts on the current matrix as the elementary row operation $R_i \leftarrow R_i \oplus R_j$. If the \CNOT{} gates in $C$ are the transvections $T^{(1)}, \ldots, T^{(k)}$ in temporal order, then the implemented matrix is $M = T^{(k)} \cdots T^{(1)}$, with the first-applied gate appearing as the rightmost factor. Hence any two \CNOT{} circuits implementing the same $M$ realize the same quantum unitary, even if their gate counts, depths, or gate orderings differ. Figure~\ref{fig:equiv3q} illustrates this on three qubits: the one-gate circuit $\CNOT(2{\to}0)$ and the time-ordered sequence $\CNOT(2{\to}1), \CNOT(2{\to}0), \CNOT(2{\to}1)$ both implement $M = I + e_0 e_2^{\top}$. The first maps $|001\rangle$ directly to $|101\rangle$ and the second routes $|001\rangle \mapsto |011\rangle \mapsto |111\rangle \mapsto |101\rangle$, in agreement on all computational-basis states. Replacing $C$ by any shorter \CNOT{} circuit with the same binary matrix therefore preserves the encoder exactly. The same principle is the algebraic basis of \CNOT{} synthesis: on four qubits, Figure~\ref{fig:equiv4q} gives two decompositions of the matrix
\[
M =
\begin{bmatrix}
1&0&0&1\\
0&1&0&1\\
0&0&1&1\\
0&0&0&1
\end{bmatrix}.
\]
Sequence~A uses five \CNOT{} gates and Sequence~B uses three, yet both build the same transformation from the identity, so the shorter sequence is an exact replacement for the longer one. Appendix~\ref{app:steps} records the per-step matrices for this example. Removing such redundancy while preserving the implemented matrix is the synthesis problem formalized in this work. We write $\sM(M)$ for the minimum number of transvections whose product is $M$, equivalently the distance from $M$ to the identity in the Cayley graph of $\GL(n,\Ftwo)$ generated by $\{T_{ij} : i \neq j\}$. Because every transvection is an involution (it is its own inverse, $T_{ij}^2 = I$) and the generating set is closed under transpose, with $T_{ij}^{\top} = T_{ji}$, the word metric satisfies $\sM(M) = \sM(M^{-1}) = \sM(M^{\top})$. We therefore state the optimization problem as follows: given the matrix $M$ extracted from a CG/SKG encoder, find a short transvection word for $M$.

\newcommand{\cxg}[3]{\fill (#1,#2) circle (1.7pt); \draw[thick] (#1,#2)--(#1,#3) (#1,#3) circle (3pt) (#1,{#3-0.13})--(#1,{#3+0.13}) ({#1-0.13},#3)--({#1+0.13},#3);}

\subsection{Circuit cost metrics}
\label{sec:costs}

Beyond word length, we use two hardware-sensitive costs. The \emph{two-qubit depth} of a circuit is its as-soon-as-possible (ASAP) depth: the number of layers when gates on disjoint qubit pairs may share a layer. The \emph{routed cost} on a coupling graph is the two-qubit gate count and depth after a routing pass inserts the SWAP gates the device connectivity requires.

\noindent\textbf{Two notions of depth.} Two metrics measure the parallel cost of a \CNOT{} sequence. The \emph{gate-list ASAP depth} schedules the gates in their listed order, placing each in the earliest layer where both its qubits are free; it is what most synthesis tools report, but it depends on that ordering. The \emph{commutation-aware depth} instead schedules the gates subject only to their genuine ordering constraints, since two \CNOT{}s commute unless the control of one is the target of the other (Section~\ref{sec:relayer}). The commutation-aware depth never exceeds the gate-list ASAP depth, and it no longer depends on how commuting gates are ordered, though it still depends on the synthesized sequence. The count and frontier tables (Tables~\ref{tab:main} and~\ref{tab:depth}) report gate-list depth to match prior work, while the depth table and the routed comparison (Tables~\ref{tab:certify} and~\ref{tab:routing}) report commutation-aware depth (see Section~\ref{sec:results}).

\section{The synthesizer: two-sided Hamming descent}
\label{sec:blts}

\subsection{Two-sided moves}
\label{sec:twosided}

Many prior compilers and optimizers operate directly on the gate list generated by CG/SKG encoder circuits. The synthesizer we propose instead operates on a residual matrix (the part of $M$ still to be eliminated). Let $E_{ij}=T_{ij}=I+e_i e_j^{\top}$ denote the transvection associated with $\CNOT(j{\to}i)$. A gate assigned to the end of a circuit left-multiplies the circuit matrix, whereas a gate assigned to the beginning right-multiplies it. Thus a target matrix $M$ can be reduced from both sides: the algorithm may either remove a final gate by a row operation or remove an initial gate by a column operation.

At every step the algorithm maintains a residual matrix $A$, a front sequence $R$, and a back sequence $L$. The front sequence contains gates already assigned to the beginning of the circuit, while the back sequence contains gates assigned to the end. These quantities satisfy the invariant
\[
A = L^{-1} M R^{-1},
\qquad\text{equivalently}\qquad
M = L A R .
\]
Initially $A=M$ and $L=R=I$. A back move chooses a candidate final gate and updates $A \leftarrow E_{ij} A$, the elementary row operation $R_i \leftarrow R_i \oplus R_j$ on the residual. A front move chooses a candidate initial gate and updates $A \leftarrow A E_{ij}$, the elementary column operation $C_j \leftarrow C_j \oplus C_i$. Synthesis terminates when $A=I$. At that point we have removed the residual completely, and the final circuit is the stored front sequence followed by the reverse of the stored back list, as described in Algorithm~\ref{alg:blts}.

\begin{figure}[htbp]
\centering
\includegraphics[width=0.52\textwidth]{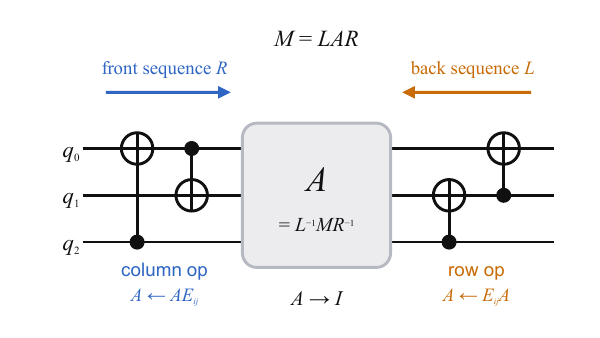}
\caption{The two-sided move space reduces the residual $A$ to the identity from both ends. Front (column-operation) and back (row-operation) moves act on $A$ while maintaining $M=LAR$. See Section~\ref{sec:twosided}.}
\label{fig:schematic}
\end{figure}

One-sided descent is obtained as the special case in which only back moves are allowed. The relevance of the two-sided space for CSS encoder synthesis is structural. The CG and SKG constructions follow fixed column-ordering choices during elimination, and these choices often leave column-shaped redundancy in the encoder matrix. Column moves can remove this structure directly, whereas a row-only descent must realize the same correction indirectly through additional row operations. Figure~\ref{fig:schematic} illustrates the two-sided construction.

\subsection{Hamming objective, depth penalty, and multistart}
\label{sec:objective}

\begin{algorithm}[t]
\caption{Two-sided Hamming descent}
\label{alg:blts}
\begin{algorithmic}[1]
\Require target $M \in \GL(n,\Ftwo)$; layer penalty $\mu$; restart count $R$
\Ensure a matrix-verified gate list, shortest over the $R$ restarts
\State $\textit{best} \leftarrow \varnothing$
\Statex \textit{// Part A: multistart restart loop under permutation conjugation}
\For{restart $r = 1, \dots, R$}
  \State draw seed permutation $P$; $A \leftarrow P M P^{\top}$; front $\leftarrow []$; back $\leftarrow []$
  \While{$A \neq I$}
    \Statex \quad\textit{// Part B: score every move by its count change $\Delta$}
    \State $D \leftarrow A \oplus I$
    \State $\Delta_R[i,j] \leftarrow \|D_i \oplus A_j\|_1 - \|D_i\|_1$, all $i \neq j$ \Comment{back gate $\mathrm{CNOT}(j{\to}i)$: $A_i \leftarrow A_i \oplus A_j$}
    \State $\Delta_C[j,i] \leftarrow \|D_{:,j} \oplus A_{:,i}\|_1 - \|D_{:,j}\|_1$, all $i \neq j$ \Comment{front gate $\mathrm{CNOT}(j{\to}i)$: $A_{:,j} \leftarrow A_{:,j} \oplus A_{:,i}$}
    \Statex \quad\textit{// Part C: add the exact per-side depth penalty $\mu$}
    \State $\mathrm{score} \leftarrow \Delta + \mu \cdot [\text{move opens a new layer on its side}]$
    \Statex \quad\textit{// Part D: commit the move (count-first fallback; strict-descent stall)}
    \If{$\min(\Delta_R, \Delta_C) \geq 0$} \textbf{fail} restart $r$ \Comment{strict-descent stall} \EndIf
    \If{$\min(\mathrm{score}) \geq 0$}
      \State apply the best raw $\Delta < 0$ move \Comment{fallback past the depth penalty}
    \Else
      \State apply the minimum-score move (random tie-break)
    \EndIf
    \State update $A$; append/prepend the gate to its side
  \EndWhile
  \Statex \textit{// Part E: assemble, relabel, and verify against $M$}
  \State $C \leftarrow$ (front $+$ reversed back), relabeled by $P^{-1}$
  \State \textbf{assert} $\mathrm{matrix}(C) = M$; \textbf{if} $|C| < |\textit{best}|$ \textbf{then} $\textit{best} \leftarrow C$ \Comment{exact-$M$ verification}
\EndFor
\State \Return $\textit{best}$
\end{algorithmic}
\end{algorithm}

Algorithm~\ref{alg:blts} is a single procedure with five visible parts, banner-labeled A through E in the pseudocode, which we walk through in order.

\emph{Part~A: multistart under permutation conjugation.} Greedy descent is landscape-sensitive, so the synthesizer restarts $R = 50$ times. In restart $r$, a seeded random permutation matrix $P$ relabels the qubits and the synthesizer rebuilds the residual $A \leftarrow P M P^{\top}$. Permutation conjugation preserves CNOT count because $P T_{ij} P^{\top} = T_{P(i),P(j)}$, so it maps transvections to transvections, and any decomposition found under the relabeling carries back to one of the same length on $M$ (Appendix~\ref{app:relabel}). This relabeling changes which row and column moves appear locally best, so different restarts follow genuinely distinct descent paths. Random tie-breaking among equal scores adds further diversity.

\emph{Part~B: residual objective and move scoring.} Inside each restart the synthesizer reduces the residual $A$ to the identity over the two-sided move space of Section~\ref{sec:twosided}. A back move is a row addition and a front move is a column addition. Each adds one \CNOT{} gate, the back move at the end of the circuit and the front move at the beginning. Progress is measured by the Hamming distance to the identity $h(A) = \|A \oplus I\|_1$, the number of entries where $A$ differs from $I$. For a candidate move, $\Delta$ is its exact effect on $h(A)$, so $\Delta < 0$ means the move brings the residual closer to the identity. The back-move table $\Delta_R$ records this change for every row addition and the front-move table $\Delta_C$ for every column addition, as written in the pseudocode. All $2n(n-1)$ moves are scored in one batch with bit-packed popcounts, costing $O(n^3/w)$ word operations per step (with machine width $w$). A single restart commits one gate per step and decreases $h(A)$ each time, so with $h(A) \le n^2$ it uses at most $n^2$ gates by strict descent and costs $O(n^5/w)$ word operations end to end.

\emph{Part~C: the layer penalty $\mu$ and the count--depth trade-off.} We track the circuit's real depth as it grows, updating it directly from the layers formed by the gates chosen so far rather than approximating it from a heuristic such as critical-path length. Because the circuit grows from both ends, each end keeps its own running ASAP layer count, tracking which qubits are occupied in the current layer and how many layers deep that end is. A move can join its side's current layer when neither of its qubits is already used there, adding a \CNOT{} gate without increasing depth. If it instead needs a new layer, we add a penalty $\mu$ to its score, giving $\mathrm{score} = \Delta + \mu \cdot [\text{move opens a new layer on its side}]$, where $\Delta$ is the change in the Hamming objective $h(A)$. Setting $\mu = 0$ optimizes count alone, while larger $\mu$ prefers moves that keep the circuit shallower. Sweeping $\mu \in \{0.5, 1, 2, 4, 8, 16\}$ traces a count--depth Pareto frontier in a single framework. A small penalty can even lower the final count: on BB $[\![72,12,6]\!]$ the count improves from $297$ at $\mu = 0$ to $290$ at $\mu = 0.5$. The reason is that a small $\mu$ stops the greedy descent from piling gates onto the same qubits just to lower the count quickly, since those moves make the circuit deeper and often leave more gates to remove later. Avoiding them guides the search toward decompositions that are both shorter and more parallel.

During the descent, the per-side layer counts serve only as a running depth estimate to guide move selection. They are not the final reported depth. Because the circuit is built from both ends, we track the front and back sequences separately and schedule each sequence ASAP as it grows. This gives the exact ASAP depth of each partial sequence. When synthesis terminates, the final circuit is the front sequence followed by the reversed back sequence. Reversing a sequence does not change its layer count, so the sum of the two per-side depths is an upper bound on the gate-list depth of the final circuit. This bound can be loose, because a fresh ASAP schedule of the concatenated circuit can place gates from the reversed back sequence into layers left unused by the front sequence. For this reason, the $\mu$ penalty is only used to guide move selection during synthesis. The reported gate-list depth is always recomputed from the final concatenated circuit, while the commutation-aware depth is computed separately by the scheduling pass in Section~\ref{sec:relayer}.

\emph{Part~D: move selection and stall handling.} The restart continues only while there is a move that strictly decreases the unpenalized Hamming objective. If all entries of $\Delta_R$ and $\Delta_C$ are nonnegative, no row or column transvection makes progress toward the identity. The restart is then declared stalled. Otherwise, the synthesizer commits one move. If the layer penalty makes every penalized score nonnegative, the algorithm ignores the penalty for that step and applies the best move with $\Delta < 0$. If at least one penalized score is negative, it applies a minimum-score move, breaking ties at random. Thus the depth penalty guides the descent, but it does not interfere with Hamming-distance progress.

\emph{Part~E: assembly, relabeling, and verification.} When the residual reaches $A = I$, the descent is complete. The final circuit is obtained by concatenating the stored front sequence with the reversed back sequence. We then relabel the gates by $P^{-1}$, which maps the restart labeling back to the original qubit labels. We then recompute the binary matrix implemented by the final gate list. If this matrix is not exactly $M$, the run is discarded. Among the runs that pass this equality check, the synthesizer keeps the shortest circuit. This circuit is reproducible from its $(\mu,\mathrm{seed})$ pair. In our experiments, every code and every value of $\mu$ produced at least one exact decomposition of $M$. For $\mu=0$, twelve of the seventeen codes converged in all fifty restarts. The densest EA instance was the hardest case: at $\mu=0$ only five of fifty restarts converged, and those runs produced a valid $203$-gate decomposition; the $201$-gate circuit reported in Table~\ref{tab:main} for this code comes from the $\mu=0.5$ sweep.

Prior greedy methods~\cite{schaeffer2014cost,de2021gaussian} score each move by how much it reduces the matrix weight, the number of 1s it contains, using objectives such as hsum, hprod, and the variants Hsum and Hprod. Because every permutation matrix minimizes these scores equally, the descent has no reason to prefer the identity and can stall at a different permutation. In contrast, our synthesizer uses distance to the identity as the descent objective, combines it with a soft layer penalty, and explores multiple equivalent labelings through permutation-conjugation multistart. Section~\ref{sec:headtohead} evaluates these changes directly on LDPC encoder matrices.

\subsection{Commutation-aware scheduling and verification}
\label{sec:relayer}
The layer penalty $\mu$ in Section~\ref{sec:objective} influences which moves are chosen during synthesis, but it does not define the final depth of the circuit. The depth of a fixed \CNOT{} list depends on the order in which the gates are scheduled. Since many \CNOT{} gates commute, the gate-list ASAP depth can overestimate the parallel cost of the synthesized circuit. We therefore apply a separate commutation-aware scheduling pass after synthesis.

For two gates $\CNOT(c_1\to t_1)$ and $\CNOT(c_2\to t_2)$, noncommutation arises only when the control of one gate is the target of the other, that is, when $c_1=t_2$ or $c_2=t_1$. If neither condition holds, the two gates may be reordered without changing the implemented linear transformation. Thus, for a \CNOT{} sequence, the only ordering constraints that must be preserved are the precedence relations between noncommuting pairs. Gates that commute may be moved across one another, subject to the hardware rule that a qubit participates in at most one two-qubit gate in a layer.

Algorithm~\ref{alg:relayer} implements this scheduling pass. Conceptually, it builds a dependency graph whose vertices are the \CNOT{} gates and whose directed edges point from an earlier gate to a later noncommuting gate. In the implementation, this graph is not constructed explicitly. Instead, the scheduler maintains two arrays. The value $\mathrm{lc}[q]$ stores the latest layer containing a gate that used qubit $q$ as a control, and $\mathrm{lt}[q]$ stores the latest layer containing a gate that used $q$ as a target. For a new gate $\CNOT(c\to t)$, all earlier noncommuting predecessors are captured by the layer $1+\max\{\mathrm{lc}[t],\,\mathrm{lt}[c]\}$. Starting from this layer, the scheduler places the gate in the earliest layer in which both $c$ and $t$ are free. It then updates $\mathrm{lc}[c]$ and $\mathrm{lt}[t]$. This gives a near-linear-time greedy ASAP schedule of the synthesized sequence: the per-gate precedence bound is $O(1)$, and the busy-layer scan adds at most $O(d)$ per gate for output depth $d$.

\begin{algorithm}[t]
\caption{Commutation-aware scheduling and verification}
\label{alg:relayer}
\begin{algorithmic}[1]
\Require gate list $C = (g_1, \dots, g_m)$ with $\mathrm{matrix}(C) = M$
\Ensure layer map $\ell(\cdot)$ and commutation-aware depth $d$; reordered $C'$ equal to $C$
\Statex \textit{// Part A: per-qubit precedence indices (no explicit graph)}
\State $\mathrm{lc}[q] \leftarrow 0$ and $\mathrm{lt}[q] \leftarrow 0$ for all qubits $q$ \Comment{last layer $q$ is used as a control, as a target}
\Statex \textit{// Part B: greedy ASAP layering, one gate per qubit per layer}
\State mark every qubit free in every layer
\For{each gate $g_k = \CNOT(c_k \to t_k)$ in gate-list order}
  \State $\ell \leftarrow 1 + \max(\mathrm{lc}[t_k],\ \mathrm{lt}[c_k])$ \Comment{after all noncommuting predecessors, in $O(1)$}
  \While{$c_k$ or $t_k$ is busy in layer $\ell$}
    \State $\ell \leftarrow \ell + 1$ \Comment{both qubits must be free in the layer}
  \EndWhile
  \State $\ell(g_k) \leftarrow \ell$;\ \ mark $c_k, t_k$ busy in layer $\ell$;\ \ $\mathrm{lc}[c_k] \leftarrow \max(\mathrm{lc}[c_k], \ell)$;\ \ $\mathrm{lt}[t_k] \leftarrow \max(\mathrm{lt}[t_k], \ell)$
\EndFor
\State $d \leftarrow \max_k \ell(g_k)$ \Comment{commutation-aware depth}
\Statex \textit{// Part C: verify the reordering preserves the unitary}
\State $C' \leftarrow$ gates of $C$ sorted by $(\ell(g_k), k)$
\State \textbf{assert} $\mathrm{matrix}(C') = M$ \Comment{exact-$M$ matrix check}
\State \textbf{assert} $\mathrm{tableau}(C') = \mathrm{tableau}(C)$ \Comment{Clifford-tableau equality at zero \CNOT{} cost}
\State \Return $(C', d)$
\end{algorithmic}
\end{algorithm}

The resulting layer order is used only after verification. We sort the gates by their assigned layers, recompute the matrix of the reordered circuit over $\Ftwo$, and require that it equals the original encoder matrix $M$. We also check Clifford-tableau equality between the original and reordered circuits. These checks ensure that scheduling changes only the timing of the gates, not the implemented encoder. The pass therefore reduces depth at zero \CNOT{} cost and preserves exact equality with the target encoder matrix.

We do not claim that this greedy schedule is globally depth-optimal. Even for a fixed \CNOT{} sequence, minimizing the number of layers under commutation and one-gate-per-qubit constraints is a constrained scheduling problem~\cite{maslov2022depth,lee2026quantum}. We therefore report the depth produced by the greedy schedule. As a useful a posteriori lower bound, any schedule of a fixed sequence must have depth at least $\max(\delta,\lambda)$, where $\delta$ is the maximum number of gates incident on any qubit and $\lambda$ is the length of the longest chain of pairwise precedence constraints induced by noncommuting gates. Section~\ref{sec:certify} compares the schedules against this lower bound.

\subsection{Noise-aware selection}
\label{sec:selectmethod}
The $\mu$-sweep of Section~\ref{sec:objective} does not give only one circuit. It gives a set of circuits with different \CNOT{}-count and depth trade-offs (Section~\ref{sec:depth-results}). Some circuits use fewer gates, while others have smaller depth. The best choice depends on the target hardware. On a device where idle errors are large, a shallower circuit may be better because qubits spend less time waiting. On a device where two-qubit gate errors dominate, a lower-count circuit may be better because it uses fewer gates.

For this reason, we do not choose a circuit before routing. Instead, we first route every Pareto-frontier candidate onto the target coupling graph (Section~\ref{sec:routing-results}). We then score the routed circuits using a simple first-order hardware-noise model and select the circuit with the lowest estimated preparation cost. Routing before selection is important. Routing can change the physical two-qubit count, the depth, and the idle time of each circuit. Therefore, the best logical circuit before routing may not be the best physical circuit after routing. By selecting only after routing, the final choice reflects the actual hardware cost.

For a routed circuit $C$, we estimate the preparation error by adding the error contributions from all places where faults can occur,
\begin{equation}
\mathcal{E}(C) = \sum_{g \in C} p_g \;+\; \sum_{(q,t)\in\mathrm{idle}(C)} p_\mathrm{idle}(q,t) \;+\; \sum_{r\in\mathrm{reset}(C)} p_\mathrm{reset}(r).
\label{eq:errgeneral}
\end{equation}
These terms correspond to two-qubit gates, idle qubit-layers, and reset operations. In our experiments we use a uniform-noise model: every two-qubit gate has error rate $p_\text{2q}$, every idle qubit-layer has error rate $p_\text{idle}$, and reset is treated as ideal. Under this model the cost becomes
\begin{equation}
\mathcal{E}(C;\kappa) = N(C) + \kappa\, I_\mathrm{lr}(C), \qquad \kappa = \frac{p_\text{idle}}{p_\text{2q}}.
\label{eq:errsurrogate}
\end{equation}
Here $N(C)$ is the routed two-qubit gate count, and $I_\mathrm{lr}(C)$ is the idle exposure after live-range scheduling. Thus the score balances two costs: using more two-qubit gates and leaving qubits idle for more layers. For a routed circuit of depth $D(C)$, the live-range idle exposure is
\begin{equation}
I_\mathrm{lr}(C) = \sum_{q}\big(D(C) - f(q) + 1\big) - 2\,N(C),
\label{eq:idlelr}
\end{equation}
where $f(q)$ is the first layer in which qubit $q$ is used. This formula counts only the layers after the qubit becomes active. Since each qubit is reset just before its first use (Section~\ref{sec:lrmethod}), it does not accumulate idle error before layer $f(q)$. The subtraction $2\,N(C)$ removes the two active qubit slots used by each two-qubit gate, leaving only idle qubit-layers.

We use this scheduled idle exposure rather than the simpler ASAP estimate $n\,D(C)-2\,N(C)$. This is important because two routed circuits can have similar depth but different first-use layers. After just-in-time reset, those circuits can therefore have different idle exposure. Using $I_\mathrm{lr}(C)$ makes the selection rule match the circuit schedule that is executed, and Section~\ref{sec:noiseselect} gives a case where this schedule-aware score selects a lower-failure candidate that the ASAP-idle score misses.

The ratio $\kappa = p_\text{idle}/p_\text{2q}$ is the only hardware-dependent input. When $\kappa$ is close to zero, two-qubit gate errors dominate, so the best circuit is usually the one with the smallest routed two-qubit count. As $\kappa$ increases, idle error becomes more important, and a circuit with more gates can become better if it has much lower scheduled idle exposure.

For two routed candidates $C_a$ and $C_b$, the value of $\kappa$ where the preferred choice changes is
\begin{equation}
\kappa^\star = \frac{N(C_b)-N(C_a)}{I_\mathrm{lr}(C_a)-I_\mathrm{lr}(C_b)},
\label{eq:kappastar}
\end{equation}
when the numerator and denominator have the same sign. If this condition does not hold, then one candidate is better than the other for all $\kappa$ under the first-order score. Because routing can change gate count, depth, and idle exposure in different ways, the best routed circuit is not necessarily the logical circuit with the fewest \CNOT{} gates (Section~\ref{sec:pipelinecompare}).

Algorithm~\ref{alg:select} applies this post-routing selection rule to the full Pareto frontier. For each candidate it routes the circuit onto the target coupling graph, schedules the routed circuit, computes its live-range idle exposure, and evaluates the first-order cost in \eqref{eq:errsurrogate}. After all candidates have been scored, it returns the routed circuit with the lowest estimated preparation cost. Although the model keeps only first-order error contributions, it matches the full simulation results well. Across the Pareto frontier, its predicted costs are strongly correlated with the preparation-failure rates measured in Stim\cite{gidney2021stim}, with Pearson correlation between $0.90$ and $0.99$. We also tested a more detailed propagation-aware version of the score, in which each fault location is weighted by whether the propagated fault anticommutes with the live stabilizers and therefore changes the prepared encoded state. This refinement gives nearly the same choices as the first-order score of \eqref{eq:errsurrogate}, differing only by a few percent.

\begin{algorithm}[t]
\caption{Noise-aware post-routing selection from the count--depth Pareto frontier}
\label{alg:select}
\begin{algorithmic}[1]
\Require frontier $\mathcal{F} = \{C_1,\dots,C_F\}$; coupling graph $G$ on $n_\mathrm{phys}$ qubits; device ratio $\kappa$
\Ensure the routed frontier circuit of least first-order preparation cost
\State $best \leftarrow \varnothing$;\ \ $e^{\star} \leftarrow +\infty$
\For{each circuit $C \in \mathcal{F}$}
  \State $R \leftarrow \textsc{Route}(C, G)$ \Comment{best-of-$s$ SABRE onto the native graph (Section~\ref{sec:routing-results})}
  \State $(\ell,\, d) \leftarrow \textsc{Schedule}(R)$;\ \ $f(q) \leftarrow \min\{\ell(g): g \text{ acts on } q\}$ \Comment{ASAP layers, Algorithm~\ref{alg:relayer}}
  \State $I \leftarrow \sum_q (d - f(q) + 1) - 2\,|R|$ \Comment{idle-exposure surrogate for ranking; executed value from Algorithm~\ref{alg:liverange}}
  \State $e \leftarrow |R| + \kappa\, I$ \Comment{first-order routed cost, \eqref{eq:errsurrogate}}
  \If{$e < e^{\star}$}
    \State $e^{\star} \leftarrow e$;\ \ $best \leftarrow R$
  \EndIf
\EndFor
\State \Return $best$
\end{algorithmic}
\end{algorithm}

\subsection{Live-range scheduling}
\label{sec:lrmethod}
The commutation-aware scheduler of Section~\ref{sec:relayer} reduces the number of circuit layers. This lowers the number of layers in which idle errors can occur. However, depth alone does not determine the total idle exposure. What matters is the number of qubit-layers in which initialized qubits are present but not used. A qubit does not accumulate idle error before it is initialized. If the qubit is kept in the reset state, or if a reset is applied immediately before its first gate, then any earlier error is removed. We therefore use just-in-time initialization. Each qubit is prepared only when it is about to be used for the first time. We also schedule the first use of each qubit as late as the circuit dependencies allow. This keeps qubits out of the decohering part of the circuit for as long as possible.

Let $\ell(g)$ be the layer assigned to gate $g$, and let $d$ be the circuit depth. For each qubit $q$, define
\begin{equation}
f(q)=\min\{\ell(g): g \text{ acts on } q\},
\end{equation}
the first layer in which $q$ participates in a gate. Under just-in-time reset, qubit $q$ contributes to idle exposure only from layer $f(q)$ through layer $d$. Thus the total idle exposure is
\begin{equation}
I_\mathrm{lr}=\sum_q \big(d-f(q)+1\big)-2N,
\end{equation}
where $N$ is the number of two-qubit gates. The first term counts the live layers of all qubits after initialization. The subtraction $2N$ removes the two active qubit slots used by each two-qubit gate, leaving only idle qubit-layers. For a fixed depth $d$, minimizing $I_\mathrm{lr}$ is the same as maximizing $\sum_q f(q)$. In other words, we want each qubit's first use to occur as late as possible. Algorithm~\ref{alg:liverange} obtains this schedule by first scheduling the reversed gate list as soon as possible and then reflecting the layers. This gives an as-late-as-possible schedule at equal or lower depth (the reflected schedule inherits the depth of the ASAP schedule of the reversed gate list, which the order-sensitive greedy scheduler made equal or slightly smaller than the forward depth on every circuit we tested). The algorithm then places the reset, and the Hadamard preparation when needed, immediately before the qubit's first use. This scheduling pass does not change the gate list; it reschedules at the commutation-aware depth or lower (equal or lower on every circuit we tested). It only changes when qubits are initialized and when gates are placed within the same dependency constraints. Therefore, it can be applied both to a logical circuit and to a routed circuit after \textsc{swap} expansion. In the full pipeline, live-range scheduling is the final stage: it is applied to the routed circuit chosen by the noise-aware selection of Section~\ref{sec:selectmethod}, and because it changes neither the gate count nor the depth, it improves fidelity without affecting the selection.

\begin{algorithm}[t]
\caption{Live-range scheduling with just-in-time reset}
\label{alg:liverange}
\begin{algorithmic}[1]
\Require gate list $C$ on qubits $\{0,\dots,n-1\}$ with Hadamard set $S$ and $\mathrm{matrix}(C)=M$
\Ensure just-in-time-preparation schedule that reduces idle exposure at equal or lower commutation-aware depth
\Statex \textit{// Part A: as-late-as-possible schedule by reflecting the reversed sequence}
\State $(\ell^{R},\, d) \leftarrow \textsc{Schedule}(\mathrm{reverse}(C))$ \Comment{ASAP layers and depth, Algorithm~\ref{alg:relayer}}
\State $\ell(g) \leftarrow d + 1 - \ell^{R}(g)$ for every gate $g$ \Comment{reflect to an ALAP schedule of depth $d$}
\Statex \textit{// Part B: first-use layer and just-in-time preparation}
\For{each qubit $q \in \{0,\dots,n-1\}$}
  \State $f(q) \leftarrow \min\{\ell(g) : g \text{ acts on } q\}$ \Comment{$d{+}1$ if $q$ is unused}
  \State prepare $q$ (reset, and Hadamard if $q \in S$) in layer $f(q)-1$ \Comment{just-in-time initialization}
\EndFor
\Statex \textit{// Part C: idle exposure and verification}
\State $I \leftarrow \sum_{q} \big(d - f(q) + 1\big) - 2\,|C|$ \Comment{idle qubit-layers after first use}
\State \textbf{assert} $\mathrm{matrix}(C)=M$ \Comment{gate list and depth unchanged}
\State \Return schedule $\ell(\cdot)$, preparations, idle exposure $I$
\end{algorithmic}
\end{algorithm}

% ============================================================================
\section{Results and Analysis}
\label{sec:results}

\subsection{Setup}
\label{sec:setup}

\paragraph{Benchmarks.}
We evaluate the pipeline on seventeen CSS encoders from three families: four BB codes~\cite{bravyi2024high}, five HGP codes, and eight EA QC-LDPC codes. For each code, we first generate the reference encoder using the corresponding CG/SKG construction. We then extract its \CNOT{} block and represent it by its binary linear-reversible matrix $M$. All synthesis results are compared against this same algebraic target $M$. For the EA codes, the encoder acts on the message qubits together with the shared entanglement (ebit) qubits, so the number of qubits it spans is not directly given by the code's parameters. We therefore read each block size from the dimension of its target matrix $M$. The four BB codes are the standard $[\![72,12,6]\!]$, $[\![90,8,10]\!]$, $[\![108,8,10]\!]$, and $[\![144,12,12]\!]$ instances of Bravyi et al.~\cite{bravyi2024high}, defined over the bivariate polynomial ring $\Ftwo[x,y]/(x^{\ell}-1,\,y^{m}-1)$ by two weight-three check polynomials $A$ and $B$. The five HGP instances are built from small classical seed codes, the $[7,4,3]$ Hamming code, the length-$3$ and length-$4$ repetition codes, and two additional classical codes with parameters $[5,2]$ and $[6,3]$, leading to the HGP codes $[\![58,16]\!]$, $[\![13,1]\!]$, $[\![25,1]\!]$, $[\![34,10]\!]$, and $[\![45,9]\!]$. The eight EA instances are QC-LDPC entanglement-assisted codes from the constructions of~\cite{kumar2025entanglement,wilde2008optimal,luo2009channel}. For each code, $p$ and $\ell$ denote the circulant size and lift, and the integer after the semicolon gives the number of ebits. For example, EA $[\![25,16;1]\!]$ has $p=5$, $\ell=1$, and one ebit. The CG/SKG construction uses a fixed left-to-right column-elimination order. When a baseline circuit implements $M$ only up to a qubit permutation, we record that permutation and relabel the circuit, so every reported circuit implements the exact target matrix $M$. Logical synthesis results are reported for all seventeen codes. Routing and routed-noise experiments focus on the BB-native biplanar architecture, with HGP and EA native-routing results reported separately.

\paragraph{Synthesis protocol.}
For the main synthesis runs, we use two-sided Hamming descent with $R=50$ permutation-conjugation restarts for each $\mu \in \{0,0.5,1,2,4,8,16\}$. The reported output for a code is the best circuit from this sweep. Its \CNOT{} count is used for Table~\ref{tab:main} and its count--depth trade-off for Table~\ref{tab:depth}. To isolate the effect of the two-sided move space, we also evaluate a one-sided variant in which column moves are disabled while the objective, multistart protocol, and matrix verification remain unchanged. We test this one-sided synthesizer in two forms: a count-optimized version (the lowest count over the $\mu$-grid) and a depth-aware version. Tables~\ref{tab:main} and~\ref{tab:depth} compare against the relevant one-sided configuration.

\paragraph{Routing protocol.}
Our main routing experiments route the BB family on the BB-native biplanar coupling map of~\cite{bravyi2024high}. This is a degree-$6$ Tanner-graph architecture with $2n$ physical qubits in which the encoder acts on the $n$ data qubits and the $n$ check-plane vertices provide routing space for \textsc{swap} insertion. The HGP and EA codes are routed on their native Tanner graphs in Section~\ref{sec:native-routing}. Each logical circuit is routed using SABRE~\cite{li2019tackling} with ten random seeds at optimization level $2$. We apply the same commutation-aware re-layering pass and the same routing protocol to both the CG/SKG baseline and the synthesized circuits, so the routed comparison reflects the quality of the synthesized circuit.

\paragraph{Layout convention.}
Routing a logical circuit onto the coupling graph requires an initial layout, a mapping from logical qubits to physical qubits, and it inserts \textsc{swap} gates to bring interacting qubits together. We let SABRE choose this initial layout freely, so it can place frequently interacting qubits near one another. The inserted \textsc{swap}s permute the qubits, so the logical-to-physical mapping at the end of the circuit differs from the initial one. We accept this output permutation rather than appending further \textsc{swap}s to undo it. We use this same layout-free convention for both the CG/SKG baseline and our synthesized circuits. Thus, in the routed-depth and routed-noise tables, we do not force the logical qubits back to canonical data-qubit positions. This convention is valid when the downstream syndrome-extraction schedule follows the routed placement. In that case, the final permutation can be absorbed by relabeling. This relabeling has no gate cost. We also checked the cost of restoring the canonical data-plane positions. In this case, the restoration problem becomes token swapping on the biplanar graph. The restoration costs are comparable for the baseline and for our circuits: $177$ versus $188$, $260$ versus $259$, $330$ versus $314$, and $450$ versus $480$ swaps on the four BB codes. Even if this restoration cost is added to both circuits as a worst case, the routed two-qubit count reduction remains $43.2$--$50.5\%$. Therefore, the routed advantage is not an artifact of accepting SABRE's output permutation. The routed depth in Table~\ref{tab:routing} and the routed noise in Table~\ref{tab:extnoise} are reported in the layout-free setting. We bound the count cost of restoring the canonical data and check roles, but we do not separately report restored-layout depth or noise. These routed-depth and routed-noise results should therefore be read as layout-free preparation results. The canonical-role memory schedule is left to the architecture layer. Finally, every circuit reported below is verified to implement the target matrix $M$ exactly. All experiments are deterministic given the recorded seeds.

\subsection{\CNOT{} counts across the benchmark}
\label{sec:counts}

Table~\ref{tab:main} shows the primary logical \CNOT{}-count comparison across the seventeen-code benchmark. The column ``Base'' is the \CNOT{} count of the CG/SKG reference encoder. The column ``1-sided'' is the count-optimized one-sided variant of our synthesizer, obtained by disabling column moves and taking the best over the $\mu$-grid. The column ``Ours'' is the best circuit found by two-sided Hamming descent over the full $\mu$-grid with $R=50$ restarts. The value in parentheses is the corresponding ASAP depth. Bold entries in the ``Ours'' column mark strict improvements over the one-sided variant.

\begin{table}[htbp]
\caption{Logical \CNOT{} counts across the seventeen-code benchmark.}
\label{tab:main}
\centering
\small
\begin{tabular}{@{}lrrrr@{}}
\toprule
Code & $n$ & Base & 1-sided & Ours (depth) \\
\midrule
BB $[\![72,12,6]\!]$    &  72 &  638 &  361 & \textbf{290} (50) \\
BB $[\![90,8,10]\!]$    &  90 &  855 &  510 & \textbf{388} (54) \\
BB $[\![108,8,10]\!]$   & 108 & 1164 &  631 & \textbf{491} (51) \\
BB $[\![144,12,12]\!]$  & 144 & 2422 & 1075 & \textbf{775} (80) \\
HGP $[\![58,16]\!]$     &  58 &  183 &  157 & \textbf{149} (11) \\
HGP $[\![45,9]\!]$      &  45 &   99 &   92 & 92 (8) \\
HGP $[\![34,10]\!]$     &  34 &   72 &   69 & \textbf{68} (8) \\
HGP $[\![25,1]\!]$      &  25 &   51 &   44 & \textbf{41} (7) \\
HGP $[\![13,1]\!]$      &  13 &   20 &   19 & 19 (5) \\
EA $[\![9,4;1]\!]$      &  10 &   17 &   14 & 14 (7) \\
EA $[\![25,16;1]\!]$    &  26 &   38 &   33 & 33 (10) \\
EA $[\![25,8;1]\!]$     &  26 &   89 &   65 & \textbf{59} (14) \\
EA $[\![49,36;1]\!]$    &  50 &   82 &   73 & 73 (14) \\
EA $[\![49,12;1]\!]$    &  50 &  338 &  223 & \textbf{201} (24) \\
EA $[\![121,100;1]\!]$  & 122 &  218 &  202 & \textbf{201} (22) \\
EA $[\![6,2,2;2]\!]$    &   8 &   11 &    9 & 9 \\
EA $[\![8,2;2]\!]$      &  10 &   13 &   10 & 10 \\
\midrule
\textit{Aggregate}      &     & 6310 & 3587 & \textbf{2913} \\
\bottomrule
\end{tabular}
\end{table}

Across all seventeen encoders, two-sided Hamming descent reduces the CG/SKG total from $6310$ to $2913$ \CNOT{} gates, a $53.8\%$ aggregate reduction. The largest reductions occur on the BB family. The four BB counts decrease by $54.5\%$, $54.6\%$, $57.8\%$, and $68.0\%$, respectively. This family accounts for most of the aggregate reduction: it contributes $5079$ of the $6310$ baseline gates and falls by $61.7\%$. The sparser HGP and EA instances also improve, but by a smaller aggregate amount, from $1231$ to $969$ gates, or $21.3\%$. The routed and routed-noise results below are therefore reported on the BB codes.

The one-sided comparison shows the benefit of allowing both row and column moves. The full two-sided sweep improves ten of the seventeen instances relative to the one-sided synthesizer and ties the remaining seven. The largest gains again occur on the BB family, where the reductions relative to the one-sided variant are $19.7\%$, $23.9\%$, $22.2\%$, and $27.9\%$. Outside the BB family, the two-sided method improves EA $[\![49,12;1]\!]$ by $9.9\%$, EA $[\![25,8;1]\!]$ by $9.2\%$, HGP $[\![25,1]\!]$ by $6.8\%$, and HGP $[\![58,16]\!]$ by $5.1\%$. It also gives smaller strict improvements on HGP $[\![34,10]\!]$ and EA $[\![121,100;1]\!]$. The tied cases are not convergence failures. Several are small encoders that are already near their minimum count.

\subsection{General-purpose synthesizers and global resynthesis}
\label{sec:frameworks}

We next compare the proposed synthesizer with general-purpose synthesis and optimization tools. The goal is to test whether the reductions in Table~\ref{tab:main} can be obtained by existing compiler pipelines. We evaluate four representative methods. The Patel--Markov--Hayes (PMH) linear-reversible synthesizer~\cite{markov2008optimal} is applied directly to the matrix $M$, and we sweep its block-size parameter. Qiskit at optimization level~3~\cite{treinish2023qiskit} is applied to the CG/SKG encoder circuit. \texttt{t}$|$\texttt{ket}$\rangle$~\cite{sivarajah2021t} is applied using \textsc{FullPeephole} and \textsc{CliffordSimp}. PyZX~\cite{kissinger2019pyzx,duncan2020graph} is applied with \textsc{full\_reduce} to the full Hadamard-plus-\CNOT{} encoder. Table~\ref{tab:frameworks} reports the BB-family comparison, and Appendix~\ref{app:frameworks} presents the corresponding table for all seventeen benchmark codes.

\begin{table}[htbp]
\caption{Comparison with general-purpose synthesis and optimization tools on BB codes.}
\label{tab:frameworks}
\centering
\small
\begin{tabular}{@{}lrrrrrr@{}}
\toprule
Code & CG base & PMH~\cite{markov2008optimal} & Qiskit-O3 & \texttt{t}$|$\texttt{ket}$\rangle$~\cite{sivarajah2021t} & PyZX~\cite{kissinger2019pyzx} & Ours \\
\midrule
BB $[\![72,12,6]\!]$    &  638 &  494 &  638 &  638 &  438 & \textbf{290} \\
BB $[\![90,8,10]\!]$    &  855 &  681 &  855 &  855 &  560 & \textbf{388} \\
BB $[\![108,8,10]\!]$   & 1164 &  996 & 1164 & 1164 &  775 & \textbf{491} \\
BB $[\![144,12,12]\!]$  & 2422 & 1428 & 2422 & 2422 & 1480 & \textbf{775} \\
\bottomrule
\end{tabular}
\end{table}

No general-purpose method matches the proposed synthesizer on any BB code. PMH is the strongest linear-reversible method in this group. It reduces the CG/SKG counts, but its best circuits still use $1.70$--$2.03\times$ as many gates as ours. It removes some construction overhead without removing the column-order redundancy captured by the two-sided move space. Qiskit-O3 and \texttt{t}$|$\texttt{ket}$\rangle$ leave the CG/SKG \CNOT{} count unchanged on all four BB encoders. Their local, peephole-style passes cannot remove the global redundancy that the fixed CG/SKG construction builds into these encoder matrices.

PyZX gives the strongest general-purpose global resynthesis result in this comparison. It improves substantially over the CG/SKG construction, but its two-qubit counts remain $1.44$--$1.91\times$ larger than ours on the BB family. Since the PyZX counts include both \CNOT{} and \textsc{cz} gates, this is a favorable comparison for PyZX: it is allowed to use a more general Clifford representation, while our synthesizer remains in the pure-\CNOT{} linear-reversible setting. Even with this extra freedom, the proposed synthesizer gives the smallest two-qubit count for every BB code in Table~\ref{tab:frameworks}.

The full seventeen-code comparison in Appendix~\ref{app:frameworks} shows the same pattern. No general-purpose method beats the proposed synthesizer on any instance. The only ties occur on small and sparse codes: PyZX matches the $19$-gate result on HGP $[\![13,1]\!]$, and \texttt{t}$|$\texttt{ket}$\rangle$ matches the $9$-gate result on EA $[\![6,2,2;2]\!]$. Away from these small cases, the general-purpose tools remain above the proposed synthesizer, with the largest gaps on the dense, column-structured BB matrices. These results suggest that the gains in Table~\ref{tab:main} are not just generic compiler cleanups. They come from resynthesizing the exact linear-reversible target $M$ using a two-sided transvection objective tailored to the structure of CSS encoder matrices. Other synthesis families are not directly competitive at these block sizes: learning-based and reinforcement-learning approaches~\cite{romanello2025cnot} are designed for small or training-limited regimes, while exact and A$^{*}$-based methods~\cite{webster2025heuristic,shaik2024optimal} target small-$n$ optimal synthesis.

\subsection{Head-to-head comparison against strongest published greedy methods}
\label{sec:headtohead}

We next compare two-sided Hamming descent against implementations of the strongest published greedy baselines. The comparison uses the BB family, where the encoder matrices are dense and the weight-based greedy objectives are most likely to lose a useful descent direction. All methods are run under the same budget: $R=50$ restarts, the same machine, the same random tie-breaking rule, and the same underlying implementation. The baselines use the interleaved row/column move space with the hsum, hprod, Hsum, and Hprod cost functions of~\cite{schaeffer2014cost,de2021gaussian}. They are allowed to synthesize up to an output permutation, while our synthesizer is required to implement the exact target matrix $M$. This gives the baselines a slightly easier target than the exact $M$ our synthesizer must implement.

In Table~\ref{tab:headtohead}, each baseline entry reports the best \CNOT{} count found, with the number of converged restarts shown in parentheses. The word ``stall'' means that no restart converged under strict descent. The ``Best'' column is the best count among the four baseline variants, and the ``Gain'' column is $\mathrm{Gain}=(\mathrm{Best}-\mathrm{Ours})/\mathrm{Best}$, where ``Ours'' is the count-best circuit selected from the $\mu$-sweep with $R=50$. For the hprod family, we evaluated both the original row-weight column scoring rule and a strengthened, inverse-augmented version that converges more often. Table~\ref{tab:headtohead} reports the better value in each case, so this choice favors the baselines. Under the original hprod scoring rule, the hprod family stalls on all fifty restarts for BB $[\![72,12,6]\!]$ and BB $[\![144,12,12]\!]$. It reaches only $474$ and $635$ gates on BB $[\![90,8,10]\!]$ and BB $[\![108,8,10]\!]$, respectively.

\begin{table}[htbp]
\caption{Matched-budget comparison with published greedy baselines on BB codes.}
\label{tab:headtohead}
\centering
\small
\begin{tabular}{@{}lrrrrrrr@{}}
\toprule
 & \multicolumn{4}{c}{Baselines of~\cite{schaeffer2014cost,de2021gaussian}} & & & \\
\cmidrule(lr){2-5}
Code & hsum & Hsum & hprod & Hprod & Best & Ours & Gain \\
\midrule
BB $[\![72,12,6]\!]$   & stall & 314 (9/50) & stall & 349 (50/50) & 314  & \textbf{290} & $7.6\%$ \\
BB $[\![90,8,10]\!]$   & stall & stall       & 435 (5/50) & 474 (30/50) & 435  & \textbf{388} & $10.8\%$ \\
BB $[\![108,8,10]\!]$  & stall & stall       & stall       & 635 (40/50) & 635 & \textbf{491} & $22.7\%$ \\
BB $[\![144,12,12]\!]$ & stall & stall       & stall       & 1028 (37/50) & 1028 & \textbf{775} & $24.6\%$ \\
\bottomrule
\end{tabular}
\end{table}

The proposed synthesizer improves on the strongest baseline variant for every BB code. The reduction ranges from $7.6\%$ on BB $[\![72,12,6]\!]$ to $24.6\%$ on BB $[\![144,12,12]\!]$. The separation is the largest on the larger BB instances, which are the main dense structured matrices targeted by this work. The comparison is restricted to the BB family because it is the relevant stress test for these greedy objectives. On the denser BB matrices, the weight-based objectives often stall. On the smaller and sparser HGP and EA encoders, both methods often reach the same near-minimal count, so the head-to-head is less informative.

The advantage is not caused by the breadth of the $\mu$-sweep. At the strictly matched single configuration, with $\mu=0$ and count-only scoring, our counts are $297$, $388$, $491$, and $775$ on the four BB codes. These compare with the strongest baseline counts of $314$, $435$, $635$, and $1028$, giving a $5.4\%$ to $24.6\%$ reduction. On BB $[\![90,8,10]\!]$, BB $[\![108,8,10]\!]$, and BB $[\![144,12,12]\!]$, the matched count already equals our reported count in Table~\ref{tab:headtohead}. On BB $[\![72,12,6]\!]$, the sweep improves the matched count only from $297$ to $290$. Thus most of the gain comes from the Hamming-distance objective and the two-sided move space, not from a wider search.

The table also shows a reliability gap. Ten of the sixteen baseline configurations stall on all fifty restarts. For $n\ge 108$, only Hprod produces any converged baseline circuit. In contrast, the proposed objective produced a circuit for every code and every synthesis configuration used in Table~\ref{tab:main}, and it converges where these baselines stall even at the strictly matched single configuration. This indicates that the improvement is not only a lower final count, but also a more reliable descent direction on these matrices. The runtime results are consistent with this behavior. Producing the best baseline circuit requires running all four cost-function variants, since most stall and the surviving one is not known in advance. Table~\ref{tab:runtime} compares the proposed synthesizer at its matched single configuration against this full four-variant sweep, at $R=50$ restarts on one machine and a shared numerical core. The proposed synthesizer is $3.4$ to $10.5\times$ faster, with the largest gap on the densest code BB $[\![144,12,12]\!]$, where every baseline variant stalls while our configuration produces a $775$-gate circuit in $102$ seconds. We attribute the gap to fewer descent steps, fewer wasted stalled restarts, and integer-only scoring.

\begin{table}[htbp]
\caption{Synthesis runtime on BB codes: the proposed synthesizer at its matched single configuration versus the full four-variant baseline sweep ($R=50$ restarts, one machine, shared numerical core).}
\label{tab:runtime}
\centering
\small
\begin{tabular}{@{}lrrr@{}}
\toprule
Code & Ours (s) & Baselines, all variants (s) & Speedup \\
\midrule
BB $[\![72,12,6]\!]$    &  23.7 &   81.6 & $3.4\times$ \\
BB $[\![90,8,10]\!]$    &  15.3 &  104.1 & $6.8\times$ \\
BB $[\![108,8,10]\!]$   &  29.7 &  186.2 & $6.3\times$ \\
BB $[\![144,12,12]\!]$  & 102.1 & 1074.9 & $10.5\times$ \\
\bottomrule
\end{tabular}
\end{table}

The comparison is conservative in two ways. First, for the hprod family, we report the stronger value whenever the strengthened variant improves on the original implementation. Second, the published baselines are allowed to synthesize up to an output permutation, while our method is verified against the exact target $M$, so we compare the baselines on \CNOT{} count only. Both choices favor the baselines.

\subsection{Count--depth Pareto frontiers}
\label{sec:depth-results}

The layer-penalty sweep produces a set of circuits rather than a single output. Each value of $\mu$ can lead to a different descent path. Each point in Table~\ref{tab:depth} is a distinct circuit with its own \CNOT{} count and two-qubit depth. The parameter $\mu$ acts during synthesis, penalizing moves that open a new ASAP layer. It therefore changes the gate sequence itself, rather than only reordering the gates of a circuit that is already built.

\begin{table}[htbp]
\caption{Selected count--depth points from the one-sided, depth-greedy~\cite{de2021reducing}, and two-sided Pareto frontiers on BB codes.}
\label{tab:depth}
\centering
\small
\adjustbox{max width=\textwidth}{%
\begin{tabular}{@{}llll@{}}
\toprule
Code & One-sided baseline & Depth-greedy~\cite{de2021reducing} & Ours (selected points) \\
\midrule
BB $[\![72,12,6]\!]$   & $(361,50)$, $(374,29)$, $(402,26)$ & $(401,44)$, $(411,39)$ & $(290,50)$, $(304,28)$, $(317,24)$ \\
BB $[\![90,8,10]\!]$   & $(510,66)$, $(511,30)$, $(536,28)$ & $(544,43)$ & $(388,54)$, $(408,28)$, $(444,24)$ \\
BB $[\![108,8,10]\!]$  & $(631,50)$, $(666,32)$, $(706,30)$ & $(702,52)$, $(719,48)$ & $(491,51)$, $(494,34)$, $(524,28)$ \\
BB $[\![144,12,12]\!]$ & $(1075,61)$, $(1100,49)$, $(1157,47)$ & $(1110,67)$, $(1144,63)$ & $(775,80)$, $(790,47)$, $(814,36)$, $(845,34)$ \\
\bottomrule
\end{tabular}}
\end{table}

The two-sided frontier improves the one-sided variant on all four BB codes. At matched or lower depth, the two-sided circuits use substantially fewer \CNOT{} gates. For example, BB $[\![90,8,10]\!]$ improves from $(536,28)$ to $(408,28)$, and BB $[\![144,12,12]\!]$ improves from $(1157,47)$ to $(790,47)$. The depth-extreme points show the same trend: the two-sided sweep reaches depths $24$, $24$, $28$, and $34$, compared with $26$, $28$, $30$, and $47$ for the selected one-sided points. Thus the layer penalty is not simply buying depth by adding many gates. It exposes alternative decompositions that are both shorter and more parallel.

\begin{figure}[htbp]
\centering
\includegraphics[width=0.9\textwidth]{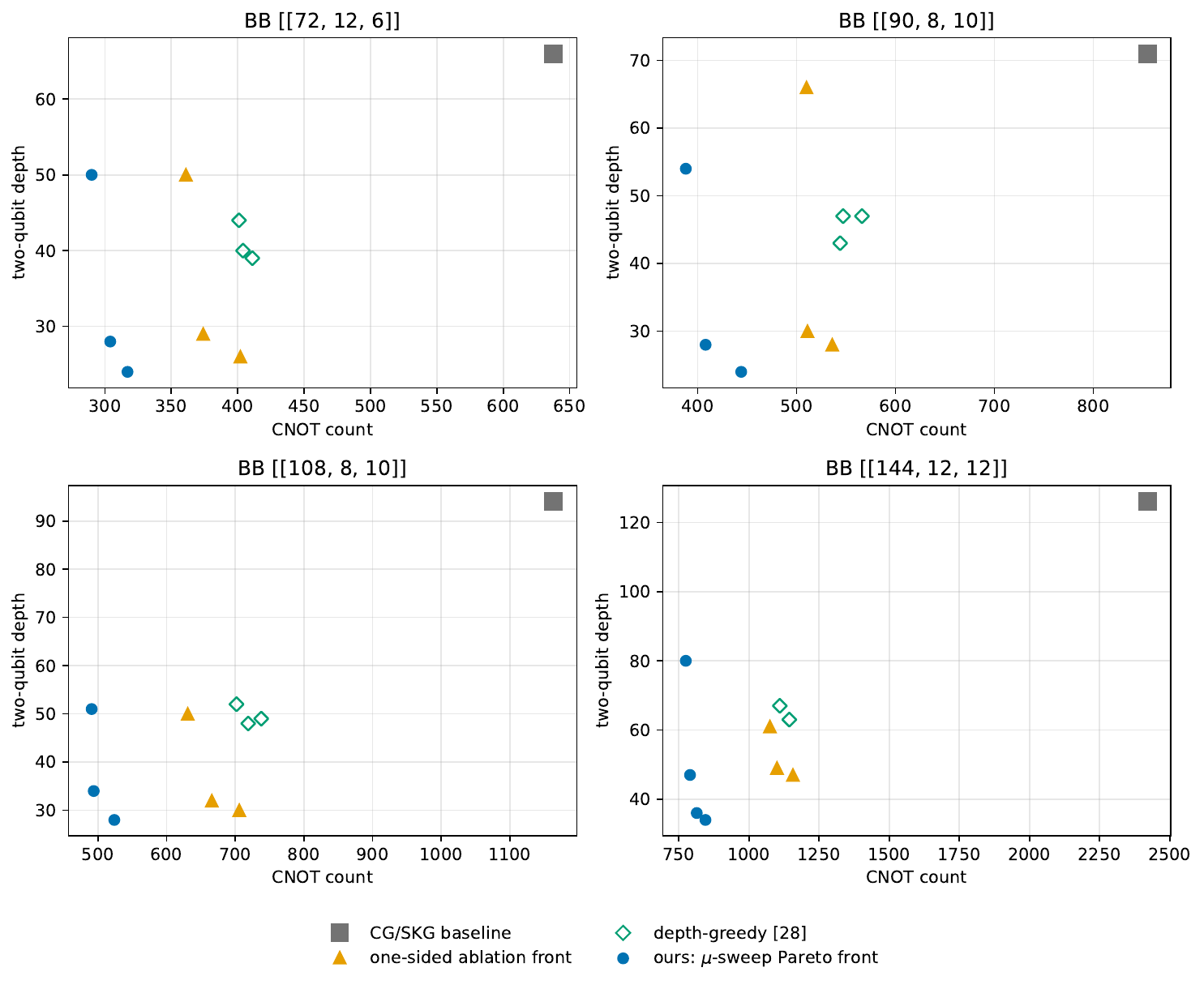}
\caption{Count--depth Pareto frontiers on the four BB codes. The two-sided front (blue, ours) dominates the one-sided ablation front (orange), the depth-greedy method of~\cite{de2021reducing} (green), and the CG/SKG construction (gray) on both count and depth for every BB code. The depth-greedy front is the Pareto-optimal subset of that method's four cost functions over many restarts.}
\label{fig:pareto}
\end{figure}

We also compare with the greedy depth method of Goubault de Brugi\`ere~\emph{et~al.}~\cite{de2021reducing}. Sweeping its four cost functions over many restarts traces a depth frontier on each BB code, shown in Figure~\ref{fig:pareto}. Our frontier dominates it on both count and depth for every code, including BB $[\![144,12,12]\!]$, where the depth-greedy method converges on only a minority of restarts. For example, our $(317,24)$ is shorter and shallower than its best point $(411,39)$ on BB $[\![72,12,6]\!]$, and our $(845,34)$ dominates its $(1110,67)$ and $(1144,63)$ on BB $[\![144,12,12]\!]$.

\subsection{Commutation-aware depth}
\label{sec:certify}
The depths in Table~\ref{tab:depth} and Figure~\ref{fig:pareto} are gate-list ASAP depths (the depth of the \CNOT{} gate list produced by the synthesizer, before any commutation-aware reordering), scheduled in the order returned by the synthesizer. That order is not unique because many \CNOT{} gates commute and can be moved without changing the implemented matrix, so we apply the commutation-aware scheduling pass of Section~\ref{sec:relayer} and verify the reordered circuit. Table~\ref{tab:certify} reports the result on the BB codes. On the count-best circuit the number of \CNOT{} gates is unchanged, while the depth decreases from $50$, $54$, $51$, and $80$ to $28$, $26$, $28$, and $31$, a $1.8$--$2.6\times$ reduction at zero \CNOT{} cost, with the implemented transformation confirmed by matrix and Clifford-tableau checks. Across the $\mu$ sweep, the lowest commutation-aware depth per code is $20$, $22$, $23$, and $27$.

\begin{table}[htbp]
\caption{Commutation-aware depth on the BB codes. For each code we show the count-best circuit (its \CNOT{} count, gate-list ASAP depth, and depth after the commutation-aware re-layering of Section~\ref{sec:relayer}) and the shallowest circuit from the $\mu$-sweep (its re-layered depth and the per-circuit lower bound $\max(\delta,\lambda)$, ratio in parentheses).}
\label{tab:certify}
\centering
\small
\begin{tabular}{@{}lccccc@{}}
\toprule
 & \multicolumn{3}{c}{Count-best circuit} & \multicolumn{2}{c}{Depth-best circuit} \\
\cmidrule(lr){2-4}\cmidrule(lr){5-6}
Code & \CNOT{}s & ASAP depth & comm.-aware & comm.-aware & lower bound \\
\midrule
BB $[\![72,12,6]\!]$   & 290 & 50 & 28 & 20 & 19 ($1.05\times$) \\
BB $[\![90,8,10]\!]$   & 388 & 54 & 26 & 22 & 19 ($1.16\times$) \\
BB $[\![108,8,10]\!]$  & 491 & 51 & 28 & 23 & 22 ($1.05\times$) \\
BB $[\![144,12,12]\!]$ & 775 & 80 & 31 & 27 & 23 ($1.17\times$) \\
\bottomrule
\end{tabular}
\end{table}

The improvement is not only relative to the original gate-list order. We also re-layer the one-sided variant under the same scheduler. Its count-best commutation-aware depths are $30$, $35$, and $36$ on BB $[\![72,12,6]\!]$, BB $[\![90,8,10]\!]$, and BB $[\![108,8,10]\!]$, against our $28$, $26$, and $28$, and on BB $[\![144,12,12]\!]$, where it converges on only a few restarts, its count-best circuit re-layers to depth $50$ against our $31$. The two-sided circuits therefore remain shallower under the commutation-aware metric, as well as under gate-list ASAP depth.

We also re-layer the depth-greedy circuits of Goubault de Brugi\`ere~\emph{et~al.}~\cite{de2021reducing} under the same scheduler, a symmetric comparison on the commutation-aware metric. Table~\ref{tab:goubdepth} reports the shallowest re-layered depth each method reaches, and ours is shallower on every code where the depth-greedy method converges. On these dense matrices the layer constraint also drives the depth-greedy method to higher \CNOT{} counts than even the count-oriented greedy of Section~\ref{sec:headtohead}, so it is dominated on count as well as depth. The same paper also gives a divide-and-conquer construction that reaches provably optimal depth, but it ignores gate count and does so only by using several times more \CNOT{}s. That trades away the low gate count that matters for encoder preparation, so we do not benchmark it here.

\begin{table}[htbp]
\caption{Shallowest commutation-aware depth on the BB codes after re-layering by the scheduler of Section~\ref{sec:relayer}: two-sided Hamming descent versus the depth-greedy method of~\cite{de2021reducing}. The ``n/c'' entry marks not converged: on BB $[\![144,12,12]\!]$ the depth-greedy method converges on too few restarts to report a shallowest-depth value.}
\label{tab:goubdepth}
\centering
\small
\begin{tabular}{@{}lcc@{}}
\toprule
Code & Ours & Depth-greedy~\cite{de2021reducing} \\
\midrule
BB $[\![72,12,6]\!]$   & $20$ & $28$ \\
BB $[\![90,8,10]\!]$   & $22$ & $32$ \\
BB $[\![108,8,10]\!]$  & $23$ & $33$ \\
BB $[\![144,12,12]\!]$ & $27$ & n/c \\
\bottomrule
\end{tabular}
\end{table}

The schedules are close to the per-circuit lower bound $\max(\delta,\lambda)$ of Section~\ref{sec:relayer}, with $\delta$ the maximum qubit gate-degree and $\lambda$ the longest noncommuting chain. On the depth-extreme circuits this bound is $19$, $19$, $22$, and $23$, so the schedules are within $1.05$--$1.17\times$ of it. We do not claim the greedy scheduler is globally optimal. The remaining gap is mainly a property of the synthesized gate list, and closing it further would require a decomposition with a smaller maximum qubit degree, which is the role of synthesis rather than scheduling. For the remaining experiments, we report commutation-aware depth as the depth metric for the synthesized and routed circuits.

\subsection{Hardware-routed results}
\label{sec:routing-results}

We next route the BB frontier circuits on the BB-native biplanar architecture of~\cite{bravyi2024high}, the degree-6 Tanner-graph coupling map with $2n$ physical qubits. Every circuit uses the same protocol, best-of-ten SABRE layout and routing at optimization level $2$, with no topology bias during synthesis.

This is a post-routing selection experiment. Each Pareto-frontier candidate and the CG/SKG baseline are re-layered by the commutation-aware scheduler and then routed, using the same pipeline for both. Any difference in routed cost therefore reflects the quality of the synthesized circuits, not the post-processing. As in the setup, we let the routed circuit end in any qubit order, since for state preparation that permutation is just a free software relabeling. These are therefore layout-free results. We route every candidate and select the best rather than biasing synthesis toward the coupling graph. We tested that alternative directly: connectivity-aware synthesis variants, from hard connectivity constraints to soft and learned distance biases, all routed to higher cost than routing-then-selection on the same candidates. The reason is mechanical. A routing \textsc{swap} is stateful, so moving a qubit closer helps every later gate, whereas a distance penalty applied during synthesis re-pays that transport cost locally at each step, without the amortization.

\begin{table}[htbp]
\caption{Routed count--depth extremes on the BB-native biplanar architecture ($2n$ physical qubits, best-of-ten SABRE at optimization level 2). Entries are (routed two-qubit count, routed two-qubit depth).}
\label{tab:routing}
\centering
\small
\begin{tabular}{@{}lllll@{}}
\toprule
Code & Base count-best & Base depth-best & Ours count-best & Ours depth-best \\
\midrule
BB $[\![72,12,6]\!]$ & $(2716,676)$ & $(3076,463)$ & $(1279,194)$ & $(1322,184)$ \\
BB $[\![90,8,10]\!]$ & $(3997,944)$ & $(4562,555)$ & $(1935,329)$ & $(2269,246)$ \\
BB $[\![108,8,10]\!]$ & $(5696,1375)$ & $(6719,883)$ & $(2560,312)$ & $(2560,312)$ \\
BB $[\![144,12,12]\!]$ & $(11312,2193)$ & $(15724,1680)$ & $(4827,528)$ & $(4986,486)$ \\
\bottomrule
\end{tabular}
\end{table}

Table~\ref{tab:routing} reports the routed Pareto extremes (count-best and depth-best) for both the baseline (CG/SKG) and ours, each entry a pair of routed two-qubit count and depth. Routing depends on gate order, so we route each candidate in its commutation-aware scheduled order rather than its raw emitted order. This carries the depth improvement through routing instead of letting the router undo it. On the count-best BB $[\![72,12,6]\!]$ circuit, re-layering before routing lowers the routed depth from $263$ to $194$, a $26\%$ drop, so the improved logical order survives routing rather than being undone by it.

Against the symmetrically routed CG/SKG baseline, our circuits reduce routed two-qubit count by $51.6$--$57.3\%$ and routed two-qubit depth by $55.7$--$71.1\%$. Our best routed depths are $184$, $246$, $312$, and $486$ layers on the four BB codes, against the re-layered baseline's $463$, $555$, $883$, and $1680$, for example $(3076,463) \to (1322,184)$ on BB $[\![72,12,6]\!]$ and $(15724,1680) \to (4986,486)$ on BB $[\![144,12,12]\!]$. Routing the full $\mu$-sweep, rather than the count-best circuit alone, supplies the depth-favorable representatives, since the depth-best routed circuit is not always the logical count-best one. On BB $[\![90,8,10]\!]$, for instance, the frontier reaches routed depth $246$ while the count-best routed circuit has depth $329$.

\subsection{Code-native routing beyond the BB family}
\label{sec:native-routing}

\begin{figure}[htbp]
\centering
\includegraphics[width=0.72\textwidth]{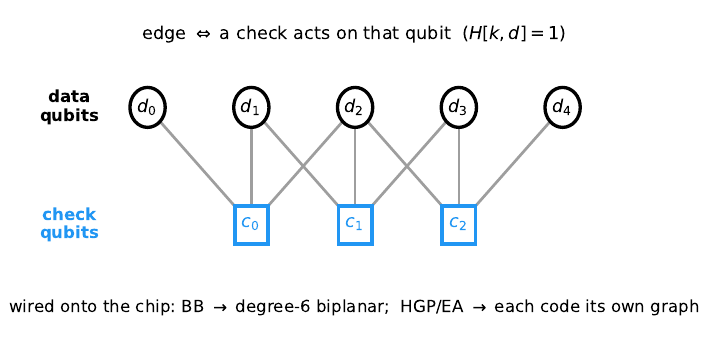}
\caption{Code-native Tanner-graph coupling map for a CSS code. Data qubits and check nodes are physical qubits, with an edge wherever $H[k,d]=1$. The check qubits provide routing space matched to the code's locality.}
\label{fig:tanner}
\end{figure}

The biplanar architecture of Section~\ref{sec:routing-results} is not a generic device. It is the BB code's own \emph{Tanner-graph coupling map}, one physical qubit per data qubit and one per check, with a coupler wherever a check acts on a qubit ($H[k,d]=1$). For the bivariate-bicycle structure this is the degree-6, thickness-2 biplanar graph of~\cite{bravyi2024high}, and Figure~\ref{fig:tanner} shows the construction. Every CSS LDPC code has such a native coupling map, built the same way from its own $H_X$ and $H_Z$. This is exactly the connectivity a device needs to measure the code's stabilizers. Routing the encoder on this graph is therefore the realistic hardware target. The HGP and EA codes do not share the BB biplanar graph, but each has its own native Tanner graph, so we route every HGP and EA encoder on its own code-native coupling map under the identical best-of-ten SABRE protocol.

\begin{table}[htbp]
\caption{Routed HGP and EA results on code-native Tanner-graph coupling maps. Entries are baseline (CG/SKG) $\to$ ours.}
\label{tab:nativerouting}
\centering
\small
\begin{tabular}{@{}lrrrr@{}}
\toprule
Code & count base$\to$ours & count $\downarrow$ & depth base$\to$best & depth $\downarrow$ \\
\midrule
HGP $[\![13,1]\!]$ & $36 \to 28$ & $22.2\%$ & $15 \to 13$ & $13.3\%$ \\
HGP $[\![25,1]\!]$ & $116 \to 101$ & $12.9\%$ & $30 \to 34$ & $-13.3\%$ \\
HGP $[\![34,10]\!]$ & $193 \to 178$ & $7.8\%$ & $50 \to 40$ & $20.0\%$ \\
HGP $[\![45,9]\!]$ & $314 \to 279$ & $11.1\%$ & $69 \to 70$ & $-1.4\%$ \\
HGP $[\![58,16]\!]$ & $618 \to 493$ & $20.2\%$ & $143 \to 116$ & $18.9\%$ \\
\midrule
EA $[\![6,2,2;2]\!]$ & $24 \to 18$ & $25.0\%$ & $20 \to 9$ & $55.0\%$ \\
EA $[\![8,2;2]\!]$ & $17 \to 11$ & $35.3\%$ & $10 \to 13$ & $-30.0\%$ \\
EA $[\![9,4;1]\!]$ & $39 \to 31$ & $20.5\%$ & $37 \to 17$ & $54.1\%$ \\
EA $[\![25,16;1]\!]$ & $74 \to 61$ & $17.6\%$ & $34 \to 25$ & $26.5\%$ \\
EA $[\![25,8;1]\!]$ & $259 \to 172$ & $33.6\%$ & $74 \to 58$ & $21.6\%$ \\
EA $[\![49,36;1]\!]$ & $130 \to 110$ & $15.4\%$ & $49 \to 31$ & $36.7\%$ \\
EA $[\![49,12;1]\!]$ & $1183 \to 749$ & $36.7\%$ & $226 \to 163$ & $27.9\%$ \\
EA $[\![121,100;1]\!]$ & $314 \to 277$ & $11.8\%$ & $95 \to 47$ & $50.5\%$ \\
\bottomrule
\end{tabular}
\end{table}

\begin{figure}[htbp]
\centering
\includegraphics[width=0.82\textwidth]{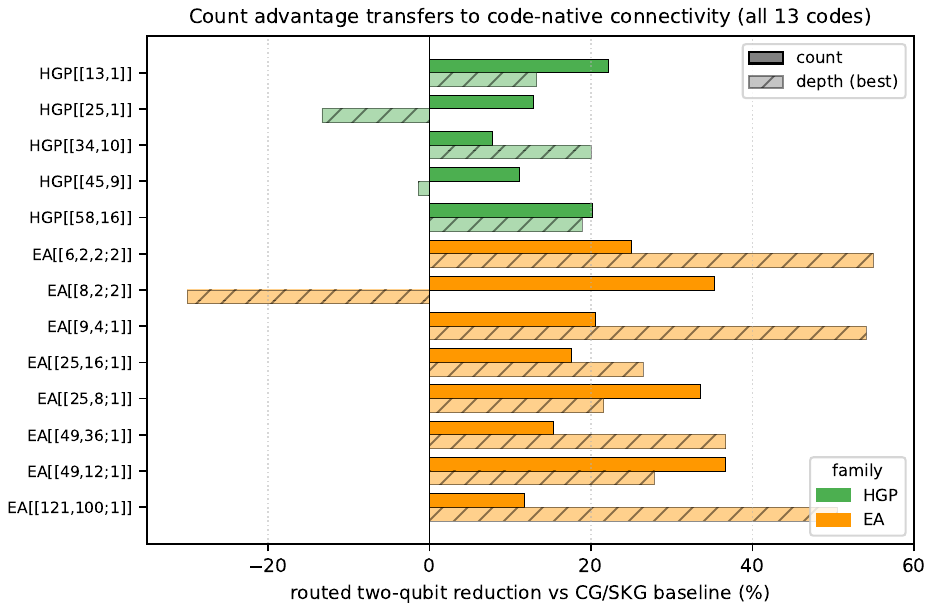}
\caption{Routed two-qubit reductions on the code-native Tanner graphs of the HGP and EA codes. Routed count falls on every code, and routed depth-best falls on ten of thirteen.}
\label{fig:native_routing}
\end{figure}

Table~\ref{tab:nativerouting} and Figure~\ref{fig:native_routing} report the result. The routed two-qubit count falls on all thirteen HGP and EA codes, by $7.8$ to $36.7\%$. Thus the count advantage transfers to code-native connectivity beyond the BB family, in the same layout-free setting used for the BB result. Taken as the better of the count-best and depth-aware circuits, the routed depth falls on ten of the thirteen codes, by up to $55\%$. On the remaining three it is neutral or at most a few layers higher, since their short routed circuits make the depth percentage sensitive to individual SWAP placements. Routing the same encoders instead on a generic degree-3 heavy-hex device compresses the count reduction, because the SABRE SWAP overhead on a sparse device unrelated to the code is a large floor common to both circuits. The code-native graph is the architecture these codes are built for, so it is the fair and favorable comparison.

\subsection{Preparation failure under circuit-level noise}
\label{sec:fidelity}

Gate count and depth are useful proxies, but the final question is whether the synthesized encoders prepare the target state more reliably on noisy hardware. We test this directly with Stim~\cite{gidney2021stim}. For each encoder, we simulate the full preparation circuit, consisting of the Hadamard layer followed by the \CNOT{} block. We apply a two-qubit depolarizing channel of strength $p$ after every \CNOT{} gate~\cite{nielsen2010quantum,fowler2012surface} and keep single-qubit gates noiseless. This isolates the noise source most directly affected by \CNOT{} synthesis. The input is the all-zero computational state. For EA codes, this input includes both the message and ebit registers. After the circuit, we measure all stabilizer generators of the ideal output state, obtained from the noiseless circuit tableau. This check is stronger than checking only the code stabilizers, because it also detects logical deviations from the intended encoded state.

For each run we measure the stabilizers of the ideal prepared state, each of which should return $+1$. If any returns $-1$, the state has deviated, so we count the run as a failure. These stabilizers determine the state completely, so the reported rate is exactly the probability of preparing a state other than the ideal state. It counts any deviation from the ideal prepared state, including errors that a later decoder might correct. It is therefore not a post-decoding logical failure rate. A $p=0$ self-test gives zero failures for every circuit. Each reported failure rate is estimated from at least $20000$ runs.

\begin{table}[htbp]
\caption{Preparation-failure rates under two-qubit depolarizing noise.  The count ratio is the baseline-to-ours \CNOT{} ratio and suppression ratio is the baseline-to-ours failure ratio. The ``ours (depth $N$)'' rows are low-depth points from the same $\mu$-sweep frontier; the two BB rows also appear in Table~\ref{tab:depth}.}
\label{tab:fidelity}
\centering
\small
\begin{tabular}{@{}llrrrr@{}}
\toprule
Code & Circuit & \CNOT{}s & $p = 10^{-3}$ & $3 \times 10^{-4}$ & $10^{-4}$ \\
\midrule
\multirow{3}{*}{BB $[\![72,12,6]\!]$} & baseline & 638 & 0.454 & 0.165 & 0.0589 \\
 & ours (count) & 290 & 0.239 & 0.080 & 0.0272 \\
 & ours (depth 28) & 304 & 0.252 & 0.084 & 0.0306 \\
 & \multicolumn{2}{l}{\emph{count / suppression ratio $2.20$}} & $1.90\times$ & $2.07\times$ & $2.17\times$ \\
\midrule
\multirow{3}{*}{BB $[\![144,12,12]\!]$} & baseline & 2422 & 0.908 & 0.515 & 0.2105 \\
 & ours (count) & 775 & 0.524 & 0.201 & 0.0729 \\
 & ours (depth 47) & 790 & 0.534 & 0.206 & 0.0779 \\
 & \multicolumn{2}{l}{\emph{count / suppression ratio $3.13$}} & $1.73\times$ & $2.57\times$ & $2.89\times$ \\
\midrule
\multirow{3}{*}{HGP $[\![58,16]\!]$} & baseline & 183 & 0.158 & 0.046 & 0.0163 \\
 & ours (count) & 149 & 0.130 & 0.038 & 0.0149 \\
 & ours (depth 9) & 150 & 0.124 & 0.042 & 0.0146 \\
 & \multicolumn{2}{l}{\emph{count / suppression ratio $1.23$}} & $1.21\times$ & $1.21\times$ & $1.10\times$ \\
\midrule
\multirow{3}{*}{EA $[\![49,12;1]\!]$} & baseline & 338 & 0.277 & 0.092 & 0.0336 \\
 & ours (count) & 201 & 0.180 & 0.059 & 0.0202 \\
 & ours (depth 23) & 206 & 0.185 & 0.059 & 0.0193 \\
 & \multicolumn{2}{l}{\emph{count / suppression ratio $1.68$}} & $1.54\times$ & $1.56\times$ & $1.67\times$ \\
\bottomrule
\end{tabular}
\end{table}

\begin{figure}[htbp]
\centering
\includegraphics[width=\textwidth]{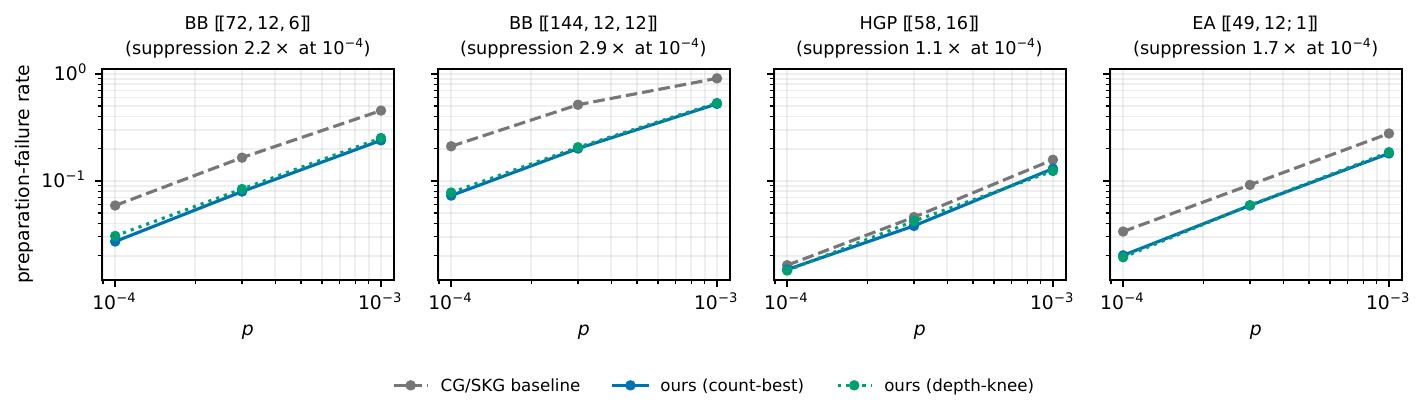}
\caption{Preparation-failure rate versus two-qubit depolarizing strength $p$ for four representative codes. The synthesized circuits outperform the CG/SKG baselines at every noise level, and the gap approaches the \CNOT{}-count ratio as $p$ decreases.}
\label{fig:fidelity}
\end{figure}

Table~\ref{tab:fidelity} reports four representative codes covering the BB, HGP, and EA families, and Figure~\ref{fig:fidelity} plots the corresponding failure rates against the two-qubit depolarizing strength $p$. The synthesized encoders prepare the correct state more often than the CG/SKG baseline in every case. The improvement is largest on the BB codes, where the \CNOT{} reductions are also largest. As $p$ decreases, the suppression approaches the \CNOT{}-count ratio. For BB $[\![72,12,6]\!]$, the suppression improves from $1.90\times$ at $p=10^{-3}$ to $2.17\times$ at $p=10^{-4}$, approaching the count ratio $2.20$. For BB $[\![144,12,12]\!]$, it improves from $1.73\times$ to $2.89\times$, approaching the count ratio $3.13$. This is the expected behavior when single-fault events dominate: at lower physical error rates, reducing the number of two-qubit gates translates almost directly into higher preparation fidelity.

The low-depth circuits have nearly the same failure rates as the count-best circuits. On the BB examples, these circuits are $41$--$44\%$ shallower than the count-best gate-list schedules, but their failure rates differ only by a few percent under this gate-error-only model. Thus the depth reduction does not noticeably hurt fidelity here. Its main benefit appears when idle errors are included, as discussed below.

The practical effect is largest on the hardest preparations. At $p=10^{-3}$, the BB $[\![144,12,12]\!]$ baseline fails on $90.8\%$ of runs, while the synthesized count-best encoder fails on $52.4\%$. Under strict post-selection on deviation-free preparation, this corresponds to about $10.9$ attempts per success for the baseline and $2.1$ for the synthesized encoder. This ratio is an upper bound on preparation overhead, since a decoder could absorb some of the correctable residual errors that this stabilizer-state test counts as failures.

The previous experiment isolates two-qubit gate noise. We also test two richer noise models. The first is a full circuit-level model~\cite{bravyi2024high}. It adds single-qubit depolarizing noise of strength $p/10$ after every Hadamard and on every idle qubit in each layer. It also adds an $X$ error of strength $p$ before readout as a measurement-error proxy. Under this model, depth becomes important because idle layers also contribute errors. Scheduling the count-best circuit at its commutation-aware depth instead of its gate-list depth lowers the BB full-model failure rate by $1.2$--$1.5\times$. For example, at $p=10^{-3}$, the rate drops from $0.47$ to $0.38$ on BB $[\![72,12,6]\!]$, and from $0.84$ to $0.69$ on BB $[\![144,12,12]\!]$. The second enrichment is a routed noise model. We use the circuits routed on the BB biplanar architecture from Section~\ref{sec:routing-results}. Each \textsc{swap} is decomposed into three \CNOT{} gates, and two-qubit depolarizing noise is applied after every physical two-qubit gate. The ideal-output stabilizers are built from the same routed circuit, so the final SABRE layout permutation is handled exactly.

\begin{table}[htbp]
\caption{Preparation-failure suppression under fuller and routed noise models at $p=10^{-4}$.}
\label{tab:extnoise}
\centering
\small
\begin{tabular}{@{}lrrr@{}}
\toprule
Code & count-only & full model & routed \\
\midrule
BB $[\![72,12,6]\!]$   & $2.17\times$ & $1.57\times$ & $1.93\times$ \\
BB $[\![144,12,12]\!]$ & $2.89\times$ & $1.92\times$ & $1.79\times$ \\
HGP $[\![58,16]\!]$    & $1.10\times$ & $1.37\times$ & \textendash \\
EA $[\![49,12;1]\!]$   & $1.67\times$ & $1.78\times$ & \textendash \\
\bottomrule
\end{tabular}
\end{table}

Table~\ref{tab:extnoise} shows that the advantage is robust. Adding idle, single-qubit, and measurement noise reduces the suppression on the BB codes because these extra faults create a common error floor for both circuits. For example, BB $[\![144,12,12]\!]$ changes from $2.89\times$ suppression in the count-only model to $1.92\times$ in the full model. The advantage is not removed, however, and the synthesized encoder still prepares the correct state more often on every reported code and noise model. The routed model gives $1.79$--$1.93\times$ suppression on the BB codes, so the fidelity gain is not only a fully connected, count-only effect. It remains visible after routing and after the output permutation is handled in the stabilizer check.

\subsection{Comparing the count-optimal and noise-aware pipelines}
\label{sec:pipelinecompare}

\begin{figure}[t]
\centering
\includegraphics[width=0.62\textwidth]{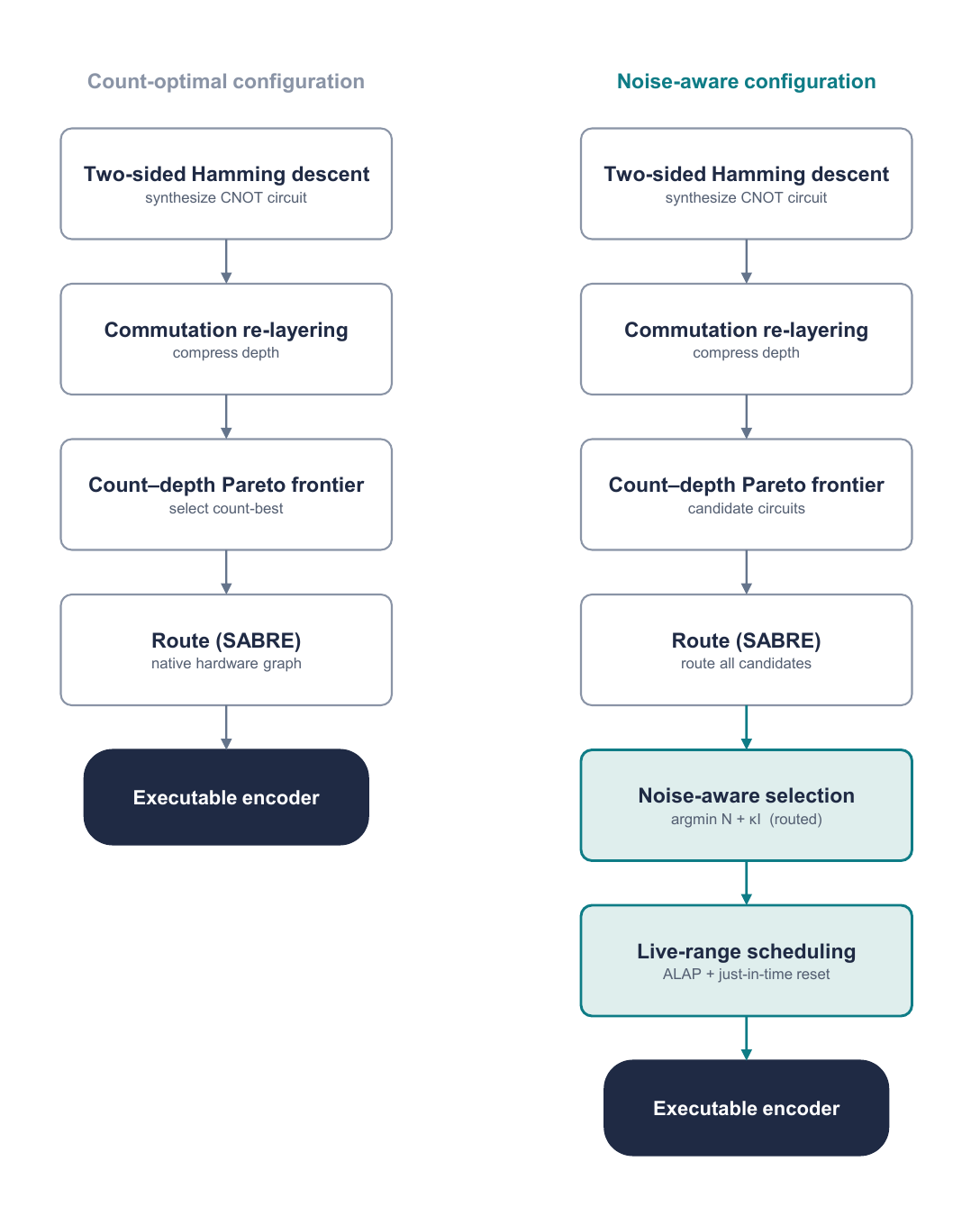}
\caption{The two pipeline configurations compared here. The count-optimal configuration selects the count-best circuit before routing. The noise-aware configuration routes the frontier, selects the lowest-cost routed candidate, and then applies live-range scheduling.}
\label{fig:pipelinecompare}
\end{figure}

We now compare two pipeline configurations. The first is a count-optimal configuration. It selects the count-best circuit, routes it, and schedules the routed circuit at its commutation-aware depth. This is the natural baseline after synthesis, routing, and re-layering. The second is the full noise-aware configuration. It routes every candidate, scores each routed circuit using the first-order cost of Algorithm~\ref{alg:select}, selects the routed circuit of lowest estimated preparation cost, and then applies live-range scheduling. The selected circuit need not be the logical count-best circuit, because routing can change both the physical \CNOT{} count and the idle structure. Live-range scheduling then changes only the timing of resets and gates. It preserves the selected routed \CNOT{} list and its two-qubit depth. Figure~\ref{fig:pipelinecompare} shows the two configurations. Table~\ref{tab:pipeline} summarizes which stages change the circuit and which use hardware or noise information.

\begin{table}[t]
\caption{Role of each pipeline stage.}
\label{tab:pipeline}
\centering
\small
\begin{tabular}{@{}llccc@{}}
\toprule
Stage & Purpose & \multicolumn{2}{c}{Changes} & Hardware input? \\
\cmidrule(lr){3-4}
 & & \CNOT{} list? & depth? & \\
\midrule
Hamming-residual resynthesis  & build shorter exact encoder       & yes                     & yes & no \\
Commutation-aware re-layering & reduce verified depth             & no                      & yes & no \\
Frontier generation           & keep count--depth candidates      & no                      & no  & no \\
Routing (SABRE)               & map to native coupling graph      & yes, via \textsc{swap}s & yes & graph \\
Noise-aware selection         & choose lowest-cost routed circuit & no                      & no  & noise model \\
Live-range scheduling         & reduce idle exposure              & no                      & no  & no \\
Stim validation               & estimate preparation failure      & no                      & no  & noise model \\
\bottomrule
\end{tabular}
\end{table}

We evaluate the full pipeline on all seventeen codes using routed Stim simulation at idle-to-gate ratio $\kappa=0.1$ and physical error rate $p=10^{-4}$. Each code is routed on its native CSS Tanner-graph coupling map, which is the biplanar graph for the BB family. The baseline is the count-best routed circuit. The full pipeline uses schedule-aware selection followed by live-range scheduling. Each reported rate is estimated from at least $200$ observed failures.

\begin{table}[t]
\caption{Preparation-failure rate of the count-best routed circuit versus the full noise-aware pipeline.}
\label{tab:ablation}
\centering
\footnotesize
\begin{tabular}{@{}lrrr@{}}
\toprule
Code & count-best routed & full pipeline & reduction \\
\midrule
BB $[\![72,12,6]\!]$   & $0.308$ & $0.243$ & $21.2\%$ \\
BB $[\![90,8,10]\!]$   & $0.440$ & $0.376$ & $14.6\%$ \\
BB $[\![108,8,10]\!]$  & $0.539$ & $0.481$ & $10.6\%$ \\
BB $[\![144,12,12]\!]$ & $0.826$ & $0.772$ & $6.5\%$  \\
\midrule
HGP $[\![13,1]\!]$     & $0.0080$ & $0.0069$ & $13.5\%$ \\
HGP $[\![25,1]\!]$     & $0.0181$ & $0.0149$ & $17.6\%$ \\
HGP $[\![34,10]\!]$    & $0.0366$ & $0.0312$ & $14.9\%$ \\
HGP $[\![45,9]\!]$     & $0.0651$ & $0.0513$ & $21.3\%$ \\
HGP $[\![58,16]\!]$    & $0.0964$ & $0.0931$ & $3.3\%$  \\
\midrule
EA $[\![6,2,2;2]\!]$   & $0.0033$ & $0.0030$ & $10.7\%$ \\
EA $[\![8,2;2]\!]$     & $0.0023$ & $0.0023$ & $-0.7\%$ \\
EA $[\![9,4;1]\!]$     & $0.0058$ & $0.0046$ & $21.6\%$ \\
EA $[\![25,16;1]\!]$   & $0.0138$ & $0.0101$ & $27.0\%$ \\
EA $[\![25,8;1]\!]$    & $0.0331$ & $0.0270$ & $18.4\%$ \\
EA $[\![49,12;1]\!]$   & $0.1414$ & $0.1206$ & $14.7\%$ \\
EA $[\![49,36;1]\!]$   & $0.0253$ & $0.0213$ & $15.8\%$ \\
EA $[\![121,100;1]\!]$ & $0.0896$ & $0.0585$ & $34.7\%$ \\
\bottomrule
\end{tabular}
\end{table}

Table~\ref{tab:ablation} shows that the full pipeline lowers preparation failure on $16$ of the $17$ codes. The largest reduction is $34.7\%$ on EA $[\![121,100;1]\!]$. The only non-improving case is EA $[\![8,2;2]\!]$, a very small circuit with only eleven routed gates, where the difference is within statistical noise. The dominant contribution comes from live-range scheduling. This stage reduces idle exposure without changing the selected routed circuit. Schedule-aware selection gives an additional benefit only when the routed frontier has multiple useful candidates. On four single-candidate instances, selection is trivial. On multi-candidate instances, it can matter substantially. The largest added selection benefit is $15.3\%$ on HGP $[\![45,9]\!]$.

The selected circuit is not always the logical count-best circuit. Routing can change the physical count and idle exposure in a nonmonotone way. For example, on BB $[\![144,12,12]\!]$, the logical count-best circuit routes to $5263$ \CNOT{} gates, while a slightly larger logical circuit routes to $5224$ \CNOT{} gates. This is why selection is performed after routing rather than before routing. Sections~\ref{sec:liverange} and~\ref{sec:noiseselect} separate the two noise-aware stages. Section~\ref{sec:liverange} measures the effect of live-range scheduling, and Section~\ref{sec:noiseselect} measures the effect of schedule-aware selection.

\subsection{Live-range scheduling: minimizing idle exposure}
\label{sec:liverange}

Section~\ref{sec:lrmethod} introduced live-range scheduling. The idea is to delay
each qubit's preparation until its first use. This keeps the qubit out
of the decohering part of the circuit for as long as possible. The pass does not add
gates or change the selected routed \CNOT{} list. It only changes when qubits are
initialized and how the same dependencies are scheduled.

Across the seventeen-code benchmark, live-range scheduling removes $28$--$71\%$ of the
idle exposure of the commutation-aware schedule. This is measured at identical \CNOT{}
count and equal-or-lower depth. Under the full noise model of Section~\ref{sec:fidelity},
with idle and measurement errors and idle-to-gate ratio $0.1$, this reduces preparation
failure on $16$ of the $17$ codes. The largest reduction from this stage alone is
$22.9\%$. On the four BB codes, the reduction is $5.9$--$8.5\%$. The largest gains occur on sparse EA codes. These circuits often bring many qubits into
use only late in the preparation, so just-in-time reset removes a large amount of pre-use
idle exposure. For example, live-range scheduling improves EA $[\![121,100;1]\!]$ by
$22.9\%$ and EA $[\![25,16;1]\!]$ by $17.2\%$.

On BB $[\![72,12,6]\!]$, the live-range schedule has the same depth as the
commutation-aware schedule. This gives a fixed-count and fixed-depth test of the
idle-exposure effect. In a $20000$-run simulation, the preparation-failure rate falls
from $0.388$ to $0.343$ at idle ratio $0.1$, a $11.6\%$ reduction with $9.4\sigma$
significance.\footnote{Significance is the two-proportion $z$-score $d/\mathrm{SE}$, where $d$ is the difference in estimated failure rates and $\mathrm{SE}=\sqrt{p_1(1-p_1)/N+p_2(1-p_2)/N}$ over $N$ runs.} At idle ratio $0.2$, it falls from $0.457$ to $0.409$, a $10.4\%$ reduction with $9.6\sigma$ significance. The benefit becomes more important after routing. Routing inserts \textsc{swap} chains that often leave many qubits waiting while others are moved. Thus
routing increases idle exposure more strongly than it increases the logical gate count.
We test this with routed Stim simulations at $p=10^{-4}$ and idle-to-gate ratio $0.1$.
Each \textsc{swap} is expanded into three \CNOT{} gates. The ideal-output stabilizers are built from the routed circuit so that the final SABRE placement is handled exactly.

\begin{table}[htbp]
\caption{Effect of live-range scheduling on routed BB preparation failure.}
\label{tab:liverange}
\centering
\small
\begin{tabular}{@{}lrrrrr@{}}
\toprule
Code & routed \CNOT{}s & depth & commutation & live-range & reduction \\
\midrule
BB $[\![72,12,6]\!]$   & $1421$ & $225\!\to\!221$ & $0.307$ & $0.268$ & $12.9\%$ ($8.7\sigma$) \\
BB $[\![90,8,10]\!]$   & $2134$ & $297\!\to\!288$ & $0.436$ & $0.376$ & $13.7\%$ ($12.2\sigma$) \\
BB $[\![108,8,10]\!]$  & $2903$ & $319\!\to\!306$ & $0.527$ & $0.481$ & $8.8\%$ ($9.3\sigma$) \\
BB $[\![144,12,12]\!]$ & $5162$ & $557\!\to\!527$ & $0.798$ & $0.747$ & $6.4\%$ ($12.2\sigma$) \\
\bottomrule
\end{tabular}
\end{table}

Table~\ref{tab:liverange} shows that live-range scheduling lowers routed preparation
failure on all four BB codes. The reductions range from $6.4\%$ to $13.7\%$, and every
point is significant at $8.7\sigma$ or higher. These improvements are obtained after routing, as the final pass of the pipeline. We apply live-range scheduling after routing rather than before routing. A schedule
computed on the logical circuit can be disrupted by \textsc{swap} insertion. In contrast,
once the circuit has been routed, the first physical interaction of each qubit is fixed up
to later delays. Computing the reset timing after routing therefore captures the full idle-exposure benefit without requiring a connectivity-aware router. The
remaining routed overhead is mostly the \textsc{swap} overhead, which multiplies the
\CNOT{} count by $4.9$--$6.6\times$ on these BB codes. This is a placement and routing
cost, separate from scheduling.

Finally, live-range scheduling assumes that just-in-time reset is sufficiently reliable.
We test this by adding an $X$ error of rate $p_\text{reset}$ to each reset at the first-use
layer, against a baseline with perfect initial reset, and sweep $p_\text{reset} \in \{0,
10^{-5}, 10^{-4}, 3{\times}10^{-4}, 10^{-3}, 3{\times}10^{-3}, 10^{-2}\}$ at $p=10^{-3}$ and
idle ratio $0.1$. The advantage persists through $p_\text{reset}=10^{-3}$ on all four BB codes, so the break-even reset error $p_\text{reset}^{\star}$ lies between $10^{-3}$ and $3{\times}10^{-3}$. At $p_\text{reset}=10^{-3}$,
live-range scheduling still improves over the perfect-initial-reset baseline by
$1.8$--$3.2\%$, and at $p_\text{reset}=10^{-4}$ the improvement is $7$--$10\%$. The pass becomes harmful by $3{\times}10^{-3}$, where reset errors outweigh the idle savings. Thus the benefit is real for high-fidelity reset and correctly disappears when
reset is too noisy.

\subsection{Noise-aware selection from the count--depth Pareto frontier}
\label{sec:noiseselect}

We next test the post-routing selection rule of Section~\ref{sec:selectmethod}. For each BB
code, we route every candidate on the count--depth frontier and score the routed circuit
using the first-order cost $\mathcal{E}(C;\kappa)$ of~\eqref{eq:errsurrogate}
(Algorithm~\ref{alg:select}). The parameter $\kappa$ is the idle-to-gate error ratio.
When $\kappa=0$, the score chooses the routed circuit with the fewest physical two-qubit
gates. As $\kappa$ increases, scheduled idle exposure becomes more important, and the selected point can move to a different part of the frontier. Table~\ref{tab:kappasweep} shows the selected logical circuit as $\kappa$ is swept over
$\{0, 0.01, 0.03, 0.1, 0.3, 1.0\}$. Each entry is the logical pair of \CNOT{} count and
depth for the candidate selected after routing and scoring.

\begin{table}[t]
\caption{Frontier circuit selected by the routed first-order cost as the idle-to-gate
ratio $\kappa$ varies, on the BB biplanar architecture.}
\label{tab:kappasweep}
\centering
\small
\begin{tabular}{@{}lcccccc@{}}
\toprule
Code & $\kappa{=}0$ & $0.01$ & $0.03$ & $0.1$ & $0.3$ & $1.0$ \\
\midrule
BB $[\![72,12,6]\!]$   & (290,28) & (294,24) & (294,24) & (297,22) & (297,22) & (297,22) \\
BB $[\![90,8,10]\!]$   & (394,22) & (394,22) & (394,22) & (388,26) & (450,19) & (450,19) \\
BB $[\![108,8,10]\!]$  & (491,28) & (491,28) & (491,28) & (491,28) & (491,28) & (491,28) \\
BB $[\![144,12,12]\!]$ & (781,30) & (781,30) & (781,30) & (781,30) & (781,30) & (882,27) \\
\bottomrule
\end{tabular}
\end{table}

The selected circuit is often not the logical count-best or the logical depth-best
circuit. It can be an interior frontier point. This happens because the selection is made
after routing, and routing can change the physical count and idle structure in a way that
is not monotone in the logical count. For example, on BB $[\![90,8,10]\!]$ and
BB $[\![144,12,12]\!]$, the selected candidate at $\kappa=0$ is not the logical count-best circuit. The idle term also needs to match the schedule that will be executed. We
therefore score the circuit using the live-range idle exposure $I_\mathrm{lr}$
of~\eqref{eq:errsurrogate}, rather than the simpler ASAP-idle estimate. This distinction
changes the selected circuit on BB $[\![90,8,10]\!]$. There, the schedule-aware score
lowers preparation failure from $0.381$ to $0.372$, a $2.4\%$ reduction with $3.1\sigma$
significance over $60000$ runs. On the other three BB codes, the schedule-aware and
simpler scores select the same candidate.

Thus, noise-aware selection is a free post-processing step over the frontier. It does not add gates. It only chooses which routed candidate is passed to the
live-range schedule of Section~\ref{sec:liverange}.

\section{Discussion and Conclusion}
\label{sec:discussion-conclusion}

Our results rest on two ideas: resynthesizing the encoder into a shorter exact circuit, and then adapting that circuit to the target hardware with noise-aware selection and live-range scheduling. We begin with the resynthesis. Standard CSS encoder constructions such as CG and SKG follow fixed algebraic elimination paths. These paths guarantee a valid encoder, but they can leave substantial redundant \CNOT{} structure in the linear-reversible block, which is especially visible on the dense, structured encoder matrices produced by quantum LDPC codes. We treated encoder preparation as an exact matrix-resynthesis problem. Two-sided Hamming descent helps because it removes this redundancy from both ends of the residual matrix: a row-only descent must correct column-shaped structure indirectly through row operations, whereas the two-sided move space also allows column operations and attacks the structure left by a fixed column-elimination order directly. This is why the gains are largest on the BB codes and the denser EA instances, while small encoders already near their minimum mostly tie the one-sided baseline. The distance-to-identity objective matters for the same reason. Weight-based greedy objectives become flat on dense encoder matrices because many moves change the matrix weight by nearly the same amount, whereas $\|A\oplus I\|_1$ keeps a direct measure of progress toward the target identity residual and gives the search a useful descent direction on these structured matrices.

Across the seventeen-code benchmark, the resulting pipeline reduces the aggregate \CNOT{} count by $53.8\%$ relative to CG/SKG, with reductions of $54.5$--$68.0\%$ on the BB family. It also beats implementations of the strongest published greedy baselines on every BB code by $7.6$--$24.6\%$. These circuits are exact replacements: every reported logical circuit is verified by recomputing the matrix it implements and checking equality with the target encoder transformation $M$. The later stages turn these logical improvements into hardware-relevant ones. Commutation-aware re-layering removes order-dependent depth at zero \CNOT{} cost, reaching within $1.05$--$1.17\times$ of a per-circuit lower bound. Routing all candidates on the BB-native biplanar architecture reduces routed two-qubit count by $51.6$--$57.3\%$ and routed two-qubit depth by $55.7$--$71.1\%$. The noise-aware stages then select the routed frontier point that best matches the device error balance and apply live-range scheduling to reduce idle exposure, lowering routed preparation failure on $16$ of the $17$ benchmark codes by up to $34.7\%$ without changing the selected routed \CNOT{} list.

Overall, the results show that encoder-matrix resynthesis is an effective compiler-level tool for quantum LDPC state preparation. By combining two-sided search, commutation-aware scheduling, post-routing hardware adaptation, and live-range scheduling, the pipeline produces smaller, shallower, and lower-noise encoder circuits while preserving the target encoded state exactly.

\appendix
\section{General-purpose synthesizers across the full benchmark}
\label{app:frameworks}

Table~\ref{tab:frameworks-full} extends Table~\ref{tab:frameworks} to all
seventeen codes. PMH is the best over its block-size sweep; Qiskit -O3 and
\texttt{t}$|$\texttt{ket}$\rangle$ act on the CG/SKG encoder; PyZX entries are
total two-qubit gates. No method beats our synthesizer on any code; the only ties are
on the two smallest, sparsest instances (PyZX matches our $19$ on HGP
$[\![13,1]\!]$, and \texttt{t}$|$\texttt{ket}$\rangle$ matches our $9$-gate
result on EA $[\![6,2,2;2]\!]$). Elsewhere the methods cluster within a few gates of
one another and above ours; the separation is a dense-encoder phenomenon.

\begin{table}[htbp]
\caption{No general-purpose synthesizer beats our synthesizer on any of the seventeen codes. All linear circuits are matrix-verified to exact $M$; PyZX (Clifford) is verified by stabilizer tableau, with total two-qubit gates reported.}
\label{tab:frameworks-full}
\centering
\small
\begin{tabular}{@{}lrrrrrrr@{}}
\toprule
Code & $n$ & CG & PMH & -O3 & \texttt{tket} & PyZX & Ours \\
\midrule
BB $[\![72,12,6]\!]$ & 72 & 638 & 494 & 638 & 638 & 438 & \textbf{290} \\
BB $[\![90,8,10]\!]$ & 90 & 855 & 681 & 855 & 855 & 560 & \textbf{388} \\
BB $[\![108,8,10]\!]$ & 108 & 1164 & 996 & 1164 & 1164 & 775 & \textbf{491} \\
BB $[\![144,12,12]\!]$ & 144 & 2422 & 1428 & 2422 & 2422 & 1480 & \textbf{775} \\
HGP $[\![58,16]\!]$ & 58 & 183 & 153 & 183 & 183 & 153 & \textbf{149} \\
HGP $[\![45,9]\!]$ & 45 & 99 & 93 & 99 & 99 & 93 & \textbf{92} \\
HGP $[\![34,10]\!]$ & 34 & 72 & 72 & 72 & 72 & 71 & \textbf{68} \\
HGP $[\![25,1]\!]$ & 25 & 51 & 45 & 51 & 51 & 43 & \textbf{41} \\
HGP $[\![13,1]\!]$ & 13 & 20 & 20 & 20 & 20 & 19 & \textbf{19} \\
EA $[\![9,4;1]\!]$ & 10 & 17 & 16 & 17 & 16 & 16 & \textbf{14} \\
EA $[\![25,16;1]\!]$ & 26 & 38 & 36 & 38 & 37 & 37 & \textbf{33} \\
EA $[\![25,8;1]\!]$ & 26 & 89 & 78 & 89 & 89 & 69 & \textbf{59} \\
EA $[\![49,36;1]\!]$ & 50 & 82 & 78 & 82 & 81 & 79 & \textbf{73} \\
EA $[\![49,12;1]\!]$ & 50 & 338 & 274 & 338 & 336 & 243 & \textbf{201} \\
EA $[\![121,100;1]\!]$ & 122 & 218 & 210 & 218 & 217 & 210 & \textbf{201} \\
EA $[\![6,2,2;2]\!]$ & 8 & 11 & 12 & 11 & 9 & 14 & \textbf{9} \\
EA $[\![8,2;2]\!]$ & 10 & 13 & 13 & 13 & 12 & 13 & \textbf{10} \\
\bottomrule
\end{tabular}
\end{table}

\section{Relabeling invariance under permutation conjugation}
\label{app:relabel}

Each restart conjugates the target by a permutation matrix $P$ (Section~\ref{sec:objective}); we record why this changes only the descent path and never the achievable \CNOT{} count. Write $P(\cdot)$ for the induced permutation, so $Pe_k = e_{P(k)}$. Since $PP^{\top}=I$, the transvection $T_{ij}=I+e_i e_j^{\top}$ satisfies
\[
P\,T_{ij}\,P^{\top} = PP^{\top} + (Pe_i)(Pe_j)^{\top} = I + e_{P(i)}e_{P(j)}^{\top} = T_{P(i),P(j)},
\]
so conjugation relabels the two qubits of a \CNOT{} without creating or removing gates. A decomposition $M=\prod_k T_{a_k b_k}$ therefore maps to $PMP^{\top}=\prod_k T_{P(a_k),P(b_k)}$ with the same number of factors, and conversely a length-$L$ decomposition of $PMP^{\top}$ gives $M = P^{\top}(PMP^{\top})P$ of length $L$ by the same identity applied with $P^{-1}$. A circuit found for the relabeled residual thus transfers back to $M$ at the same count by undoing the relabeling, the $P^{-1}$ step in Part~E of Algorithm~\ref{alg:blts}. Restarts under different $P$ explore genuinely different greedy paths at no cost to the optimum.

\section{Step-by-step transformation of the worked example}
\label{app:steps}

For the four-qubit example of Figure~\ref{fig:equiv4q}, we record the cumulative matrix after each gate, starting from $M_0 = I_4$ and applying each $\CNOT(c{\to}t)$ as the row operation $R_t \leftarrow R_t \oplus R_c$. Bit strings are ordered $q_0 q_1 q_2 q_3$. Both sequences terminate at the same $M$.

\smallskip
\noindent\emph{Sequence A} ($5$ \CNOT{}s):
\begin{gather*}
M_0 = \binmat{1&0&0&0\\0&1&0&0\\0&0&1&0\\0&0&0&1}
\xrightarrow{R_0 \leftarrow R_0 \oplus R_3} \binmat{1&0&0&1\\0&1&0&0\\0&0&1&0\\0&0&0&1}
\xrightarrow{R_1 \leftarrow R_1 \oplus R_0} \binmat{1&0&0&1\\1&1&0&1\\0&0&1&0\\0&0&0&1}
\xrightarrow{R_2 \leftarrow R_2 \oplus R_3} \binmat{1&0&0&1\\1&1&0&1\\0&0&1&1\\0&0&0&1} \\
\xrightarrow{R_1 \leftarrow R_1 \oplus R_0} \binmat{1&0&0&1\\0&1&0&0\\0&0&1&1\\0&0&0&1}
\xrightarrow{R_1 \leftarrow R_1 \oplus R_3} \binmat{1&0&0&1\\0&1&0&1\\0&0&1&1\\0&0&0&1} = M.
\end{gather*}

\noindent\emph{Sequence B} ($3$ \CNOT{}s):
\begin{gather*}
M_0
\xrightarrow{R_0 \leftarrow R_0 \oplus R_3} \binmat{1&0&0&1\\0&1&0&0\\0&0&1&0\\0&0&0&1}
\xrightarrow{R_1 \leftarrow R_1 \oplus R_3} \binmat{1&0&0&1\\0&1&0&1\\0&0&1&0\\0&0&0&1}
\xrightarrow{R_2 \leftarrow R_2 \oplus R_3} \binmat{1&0&0&1\\0&1&0&1\\0&0&1&1\\0&0&0&1} = M.
\end{gather*}

Both reach the same $M$. In Sequence~A the two $\CNOT(0{\to}1)$ gates (steps $2$ and $4$) both add row $0$ to row $1$, with row $0$ unchanged between them since the intervening $\CNOT(3{\to}2)$ acts on the disjoint row $2$. They therefore commute through the intervening gate and cancel ($T_{10}T_{23}T_{10}=T_{23}$). Sequence~A therefore realizes $T_{13}T_{23}T_{03}$, identical to Sequence~B.

\bibliographystyle{unsrt}
\bibliography{refs}

@article{shor1995scheme,
  title={Scheme for reducing decoherence in quantum computer memory},
  author={Shor, Peter W},
  journal={Physical review A},
  volume={52},
  number={4},
  pages={R2493},
  year={1995},
  publisher={APS}
}

@article{breuckmann2021quantum,
  title={Quantum low-density parity-check codes},
  author={Breuckmann, Nikolas P and Eberhardt, Jens Niklas},
  journal={PRX quantum},
  volume={2},
  number={4},
  pages={040101},
  year={2021},
  publisher={APS}
}

@article{roffe2019quantum,
  title={Quantum error correction: an introductory guide},
  author={Roffe, Joschka},
  journal={Contemporary Physics},
  volume={60},
  number={3},
  pages={226--245},
  year={2019},
  publisher={Taylor \& Francis}
}

@article{bravyi2024high,
  title={High-threshold and low-overhead fault-tolerant quantum memory},
  author={Bravyi, Sergey and Cross, Andrew W and Gambetta, Jay M and Maslov, Dmitri and Rall, Patrick and Yoder, Theodore J},
  journal={Nature},
  volume={627},
  number={8005},
  pages={778--782},
  year={2024},
  publisher={Nature Publishing Group UK London}
}

@article{tillich2013quantum,
  title={Quantum LDPC codes with positive rate and minimum distance proportional to the square root of the blocklength},
  author={Tillich, Jean-Pierre and Z{\'e}mor, Gilles},
  journal={IEEE Transactions on Information Theory},
  volume={60},
  number={2},
  pages={1193--1202},
  year={2013},
  publisher={IEEE}
}

@inproceedings{hastings2021fiber,
  title={Fiber bundle codes: breaking the n 1/2 polylog (n) barrier for quantum ldpc codes},
  author={Hastings, Matthew B and Haah, Jeongwan and O'Donnell, Ryan},
  booktitle={Proceedings of the 53rd Annual ACM SIGACT Symposium on Theory of Computing},
  pages={1276--1288},
  year={2021}
}

@inproceedings{panteleev2022asymptotically,
  title={Asymptotically good quantum and locally testable classical LDPC codes},
  author={Panteleev, Pavel and Kalachev, Gleb},
  booktitle={Proceedings of the 54th annual ACM SIGACT symposium on theory of computing},
  pages={375--388},
  year={2022}
}

@inproceedings{leverrier2022quantum,
  title={Quantum tanner codes},
  author={Leverrier, Anthony and Z{\'e}mor, Gilles},
  booktitle={2022 IEEE 63rd Annual Symposium on Foundations of Computer Science (FOCS)},
  pages={872--883},
  year={2022},
  organization={IEEE}
}

@article{kumar2025entanglement,
  title={Entanglement-Assisted Quantum Quasi-Cyclic LDPC Codes with Transversal Logical Operators},
  author={Kumar, Pavan and Sharma, Abhi Kumar and Garani, Shayan Srinivasa},
  journal={arXiv preprint arXiv:2501.07363},
  year={2025}
}

@article{wilde2008optimal,
  title={Optimal entanglement formulas for entanglement-assisted quantum coding},
  author={Wilde, Mark M and Brun, Todd A},
  journal={Physical Review A—Atomic, Molecular, and Optical Physics},
  volume={77},
  number={6},
  pages={064302},
  year={2008},
  publisher={APS}
}

@article{luo2009channel,
  title={Channel simulation with quantum side information},
  author={Luo, Zhicheng and Devetak, Igor},
  journal={IEEE Transactions on Information Theory},
  volume={55},
  number={3},
  pages={1331--1342},
  year={2009},
  publisher={IEEE}
}

@article{steane1996error,
  title={Error correcting codes in quantum theory},
  author={Steane, Andrew M},
  journal={Physical Review Letters},
  volume={77},
  number={5},
  pages={793},
  year={1996},
  publisher={APS}
}

@article{calderbank1998quantum,
  title={Quantum error correction via codes over GF (4)},
  author={Calderbank, A Robert and Rains, Eric M and Shor, Peter M and Sloane, Neil JA},
  journal={IEEE Transactions on Information Theory},
  volume={44},
  number={4},
  pages={1369--1387},
  year={1998},
  publisher={IEEE}
}

@article{cleve1997efficient,
  title={Efficient computations of encodings for quantum error correction},
  author={Cleve, Richard and Gottesman, Daniel},
  journal={Physical Review A},
  volume={56},
  number={1},
  pages={76},
  year={1997},
  publisher={APS}
}

@article{gottesman1998heisenberg,
  title={The Heisenberg representation of quantum computers},
  author={Gottesman, Daniel},
  journal={arXiv preprint quant-ph/9807006},
  year={1998}
}

@book{gottesman1997stabilizer,
  title={Stabilizer codes and quantum error correction},
  author={Gottesman, Daniel},
  year={1997},
  publisher={California Institute of Technology}
}

@article{markov2008optimal,
  title={Optimal synthesis of linear reversible circuits},
  author={Markov, Ketan and Patel, Igor and Hayes, John},
  journal={Quantum Information and Computation},
  volume={8},
  number={3\&4},
  pages={0282--0294},
  year={2008}
}

@inproceedings{sharma2025encoding,
  title={Encoding of Entanglement-assisted Quantum Codes with Fault-tolerant Syndrome Measurements},
  author={Sharma, Abhi Kumar and Kumar, Pavan and Garani, Shayan Srinivasa},
  booktitle={GLOBECOM 2025-2025 IEEE Global Communications Conference},
  pages={2723--2728},
  year={2025},
  organization={IEEE}
}

@article{treinish2023qiskit,
  title={Qiskit/qiskit-metapackage: Qiskit 0.44. 0},
  author={Treinish, Matthew},
  journal={Zenodo},
  year={2023}
}

@article{sivarajah2021t,
  title={t| ket>: a retargetable compiler for NISQ devices},
  author={Sivarajah, Seyon and Dilkes, Silas and Cowtan, Alexander and Simmons, Will and Edgington, Alec and Duncan, Ross},
  journal={Quantum Science \& Technology},
  volume={6},
  number={1},
  pages={014003},
  year={2021},
  publisher={IOP Publishing}
}

@article{nam2018automated,
  title={Automated optimization of large quantum circuits with continuous parameters},
  author={Nam, Yunseong and Ross, Neil J and Su, Yuan and Childs, Andrew M and Maslov, Dmitri},
  journal={npj Quantum Information},
  volume={4},
  number={1},
  pages={23},
  year={2018},
  publisher={Nature Publishing Group UK London}
}

@article{kissinger2020reducing,
  title={Reducing the number of non-Clifford gates in quantum circuits},
  author={Kissinger, Aleks and Van De Wetering, John},
  journal={Physical Review A},
  volume={102},
  number={2},
  pages={022406},
  year={2020},
  publisher={APS}
}

@article{duncan2020graph,
  title={Graph-theoretic simplification of quantum circuits with the ZX-calculus},
  author={Duncan, Ross and Kissinger, Aleks and Perdrix, Simon and Van De Wetering, John},
  journal={Quantum},
  volume={4},
  pages={279},
  year={2020},
  publisher={Verein zur F{\"o}rderung des Open Access Publizierens in den Quantenwissenschaften}
}

@article{bravyi2021clifford,
  title={Clifford circuit optimization with templates and symbolic Pauli gates},
  author={Bravyi, Sergey and Shaydulin, Ruslan and Hu, Shaohan and Maslov, Dmitri},
  journal={Quantum},
  volume={5},
  pages={580},
  year={2021},
  publisher={Verein zur F{\"o}rderung des Open Access Publizierens in den Quantenwissenschaften}
}

@article{de2021gaussian,
  title={Gaussian elimination versus greedy methods for the synthesis of linear reversible circuits},
  author={De Brugi{\`e}re, Timoth{\'e}e Goubault and Baboulin, Marc and Valiron, Beno{\^\i}t and Martiel, Simon and Allouche, Cyril},
  journal={ACM Transactions on Quantum Computing},
  volume={2},
  number={3},
  pages={1--26},
  year={2021},
  publisher={ACM New York, NY}
}

@article{webster2025heuristic,
  title={Heuristic and Optimal Synthesis of CNOT and Clifford Circuits},
  author={Webster, Mark and Koutsioumpas, Stergios and Browne, Dan E},
  journal={arXiv preprint arXiv:2503.14660},
  year={2025}
}

@article{maslov2023cnot,
  title={CNOT circuits need little help to implement arbitrary Hadamard-free Clifford transformations they generate},
  author={Maslov, Dmitri and Yang, Willers},
  journal={npj Quantum Information},
  volume={9},
  number={1},
  pages={96},
  year={2023},
  publisher={Nature Publishing Group UK London}
}

@article{de2021reducing,
  title={Reducing the depth of linear reversible quantum circuits},
  author={De Brugiere, Timothee Goubault and Baboulin, Marc and Valiron, Beno{\^\i}t and Martiel, Simon and Allouche, Cyril},
  journal={IEEE Transactions on Quantum Engineering},
  volume={2},
  pages={1--22},
  year={2021},
  publisher={IEEE}
}

@inproceedings{romanello2025cnot,
  title={CNOT Minimal Circuit Synthesis: A Reinforcement Learning Approach},
  author={Romanello, Riccardo and Bosco, Daniele Lizzio and Cossio, Jacopo and Sutulovic, Dusan and Serra, Giuseppe and Piazza, Carla and Burelli, Paolo},
  booktitle={2025 IEEE International Conference on Quantum Artificial Intelligence (QAI)},
  pages={253--260},
  year={2025},
  organization={IEEE}
}

@article{kissinger2019cnot,
  title={CNOT circuit extraction for topologically-constrained quantum memories},
  author={Kissinger, Aleks and de Griend, Arianne Meijer-van},
  journal={arXiv preprint arXiv:1904.00633},
  year={2019}
}

@inproceedings{meijer2023dynamic,
  title={Dynamic qubit allocation and routing for constrained topologies by cnot circuit re-synthesis},
  author={Meijer-van de Griend, Arianne and Li, Sarah Meng},
  booktitle={International Conference on Quantum Physics and Logic},
  year={2023},
  organization={Open Publishing Association}
}

@inproceedings{schneider2023sat,
  title={A SAT encoding for optimal Clifford circuit synthesis},
  author={Schneider, Sarah and Burgholzer, Lukas and Wille, Robert},
  booktitle={Proceedings of the 28th Asia and South Pacific Design Automation Conference},
  pages={190--195},
  year={2023}
}

@article{shaik2024optimal,
  title={Optimal layout-aware CNOT circuit synthesis with qubit permutation},
  author={Shaik, Irfansha and van de Pol, Jaco},
  journal={arXiv preprint arXiv:2408.04349},
  year={2024}
}

@inproceedings{christensen2025exact,
  title={On exact sizes of minimal CNOT circuits},
  author={Christensen, Jens Emil and J{\o}rgensen, S{\o}ren Fuglede and Pavlogiannis, Andreas and van de Pol, Jaco},
  booktitle={International Conference on Reversible Computation},
  pages={71--88},
  year={2025},
  organization={Springer}
}

@article{amy2019controlled,
  title={On the controlled-NOT complexity of controlled-NOT--phase circuits},
  author={Amy, Matthew and Azimzadeh, Parsiad and Mosca, Michele},
  journal={Quantum Science and Technology},
  volume={4},
  number={1},
  pages={015002},
  year={2019},
  publisher={IOP Publishing}
}

@article{gidney2021stim,
  title={Stim: a fast stabilizer circuit simulator},
  author={Gidney, Craig},
  journal={Quantum},
  volume={5},
  pages={497},
  year={2021},
  publisher={Verein zur F{\"o}rderung des Open Access Publizierens in den Quantenwissenschaften}
}

@inproceedings{li2019tackling,
  title={Tackling the qubit mapping problem for NISQ-era quantum devices},
  author={Li, Gushu and Ding, Yufei and Xie, Yuan},
  booktitle={Proceedings of the twenty-fourth international conference on architectural support for programming languages and operating systems},
  pages={1001--1014},
  year={2019}
}

@article{kissinger2019pyzx,
  title={PyZX: Large scale automated diagrammatic reasoning},
  author={Kissinger, Aleks and Van De Wetering, John},
  journal={arXiv preprint arXiv:1904.04735},
  year={2019}
}

@article{nash2020quantum,
  title={Quantum circuit optimizations for NISQ architectures},
  author={Nash, Beatrice and Gheorghiu, Vlad and Mosca, Michele},
  journal={Quantum Science and Technology},
  volume={5},
  number={2},
  pages={025010},
  year={2020},
  publisher={IOP Publishing}
}

@article{pointing2024optimizing,
  title={Quanto: Optimizing quantum circuits with automatic generation of circuit identities},
  author={Pointing, Jessica and Padon, Oded and Jia, Zhihao and Ma, Henry and Hirth, Auguste and Palsberg, Jens and Aiken, Alex},
  journal={Quantum Science and Technology},
  volume={9},
  number={4},
  pages={045009},
  year={2024},
  publisher={IOP Publishing}
}

@article{sodhani2026optimizing,
  title={Optimizing Encoder Circuits of Entanglement-Assisted Quantum LDPC Codes via Beam Search},
  author={Sodhani, Aditya and Kumar, Pavan and Garani, Shayan Srinivasa and Parhi, Keshab K},
  journal={arXiv preprint arXiv:2606.11468},
  year={2026}
}

@article{sodhani2026encoder,
  title={Encoder Circuit Optimization for Non-Binary Quantum Error Correction Codes in Prime Dimensions: An Algorithmic Framework},
  author={Sodhani, Aditya and Parhi, Keshab K},
  journal={IEEE Transactions on Quantum Engineering},
  year={2026},
  publisher={IEEE}
}

@article{schaeffer2014cost,
  title={A cost minimization approach to synthesis of linear reversible circuits},
  author={Schaeffer, Ben and Perkowski, Marek},
  journal={arXiv preprint arXiv:1407.0070},
  year={2014}
}

@article{amy2013meet,
  title={A meet-in-the-middle algorithm for fast synthesis of depth-optimal quantum circuits},
  author={Amy, Matthew and Maslov, Dmitri and Mosca, Michele and Roetteler, Martin},
  journal={IEEE Transactions on Computer-Aided Design of Integrated Circuits and Systems},
  volume={32},
  number={6},
  pages={818--830},
  year={2013},
  publisher={IEEE}
}

@article{maslov2022depth,
  title={Depth optimization of CZ, CNOT, and Clifford circuits},
  author={Maslov, Dmitri and Zindorf, Ben},
  journal={IEEE Transactions on Quantum Engineering},
  volume={3},
  pages={1--8},
  year={2022},
  publisher={IEEE}
}

@article{lee2026quantum,
  title={Quantum Circuit Optimization by Graph Coloring},
  author={Lee, Hochang and Jeong, Kyung Chul and Kim, Panjin},
  journal={Quantum},
  volume={10},
  pages={1996},
  year={2026},
  publisher={Verein zur F{\"o}rderung des Open Access Publizierens in den Quantenwissenschaften}
}

@article{mondal2024optimization,
  title={Optimization of quantum circuits for stabilizer codes},
  author={Mondal, Arijit and Parhi, Keshab K},
  journal={IEEE Transactions on Circuits and Systems I: Regular Papers},
  volume={71},
  number={8},
  pages={3647--3657},
  year={2024},
  publisher={IEEE}
}

@inproceedings{mondal2024optimized,
  title={An Optimized Nearest Neighbor Compliant Quantum Circuit for 5-Qubit Code},
  author={Mondal, Arijit and Parhi, Keshab K},
  booktitle={2024 58th Asilomar Conference on Signals, Systems, and Computers},
  pages={278--282},
  year={2024},
  organization={IEEE}
}

@article{mondal2024quantum,
  title={Quantum circuits for stabilizer error correcting codes: A tutorial},
  author={Mondal, Arijit and Parhi, Keshab K},
  journal={IEEE Circuits and Systems Magazine},
  volume={24},
  number={1},
  pages={33--51},
  year={2024},
  publisher={IEEE}
}

@book{nielsen2010quantum,
  title={Quantum computation and quantum information},
  author={Nielsen, Michael A and Chuang, Isaac L},
  year={2010},
  publisher={Cambridge university press}
}

@article{fowler2012surface,
  title={Surface codes: Towards practical large-scale quantum computation},
  author={Fowler, Austin G and Mariantoni, Matteo and Martinis, John M and Cleland, Andrew N},
  journal={Physical Review A—Atomic, Molecular, and Optical Physics},
  volume={86},
  number={3},
  pages={032324},
  year={2012},
  publisher={APS}
}

\end{document}